\documentclass[manuscript]{aastex631}

\renewcommand {\thefigure}{\arabic{figure}}

\usepackage[figuresright]{rotating}
\usepackage{multirow}
\usepackage{threeparttable}
\usepackage{appendix}

\begin{document}

\title{Acceleration and Release of Solar Energetic Particles Associated with a Coronal Shock on 2021 September 28 Observed by Four Spacecraft}

\author[0000-0002-5996-0693]{Bin Zhuang}
\affiliation{Institute for the Study of Earth, Oceans, and Space, University of New Hampshire, Durham, NH, USA}

\author[0000-0002-1890-6156]{No\'{e} Lugaz}
\affiliation{Institute for the Study of Earth, Oceans, and Space, University of New Hampshire, Durham, NH, USA}

\author[0000-0002-3176-8704]{David Lario}
\affiliation{NASA, Goddard Space Flight Center, Heliophysics Science Division, Greenbelt, MD 20771, USA}

\author[0000-0002-2106-9168]{Ryun-Young Kwon}
\affiliation{Korea Astronomy and Space Science Institute, Daejeon 34055, Republic of Korea}
\affiliation{Science and Technology Policy Institute, Korea Space Policy Research Center, Sejong 30147, Republic of Korea}

\author[0000-0002-4389-5540]{Nicolina Chrysaphi}
\affiliation{Sorbonne Universit\'{e}, \'{E}cole Polytechnique, Institut Polytechnique de Paris, CNRS, Laboratoire de Physique des Plasmas (LPP), 4 Place Jussieu, 75005 Paris, France}

\author[0000-0001-6286-5809]{Jonathan Niehof}
\affiliation{Institute for the Study of Earth, Oceans, and Space, University of New Hampshire, Durham, NH, USA}

\author[0000-0003-0510-3175]{Tingyu Gou}
\affiliation{Center for Astrophysics $|$ Harvard \& Smithsonian, Cambridge, MA 02138, USA}

\author[0000-0003-3936-5288]{Lulu Zhao}
\affiliation{Department of Climate and Space Sciences and Engineering (CLaSP), University of Michigan, Ann Arbor, MI, USA}

\begin{abstract}
The main driver of the acceleration of solar energetic particles (SEPs) is believed to be shocks driven by coronal mass ejections (CMEs). 
Extreme ultraviolet (EUV) waves are thought to be the propagating footprint of the shock on the solar surface.
One of the key questions in SEP research is the timing of the SEP release with respect to the time when the EUV wave magnetically connects with an observer. Taking advantage of close-to-the-Sun measurements by Parker Solar Probe (PSP) and Solar Orbiter (SolO), we investigate an SEP event that occurred on 2021 September 28 and was observed at different locations by SolO, PSP, STEREO-A, and near-Earth spacecraft. During this time, SolO, PSP and STEREO-A shared similar nominal magnetic footpoints relative to the SEP source region but were at different heliocentric distances. We find that the SEP release times estimated at these four locations were delayed compared to the times when the EUV wave intercepted the footpoints of the nominal magnetic fields connecting to each spacecraft by around 30 to 60 minutes. Combining observations in multiple wavelengths of radio, white-light, and EUV, with a geometrical shock model, we analyze the associated shock properties, and discuss the acceleration and delayed release processes of SEPs in this event as well as the accuracy and limitations of using EUV waves to determine the SEP acceleration and release times.
\end{abstract}


\setcounter{figure}{0}

\section{Introduction}\label{sec: intro}
Solar energetic particles (SEPs) are accelerated near the Sun in association with two mechanisms, namely (a) magnetic reconnection processes in association with solar flares and jets and (b) shocks (and compressions) driven by fast coronal mass ejections (CMEs) \citep{reames1995,reames1999}. Large and gradual SEP events associated with fast and wide CMEs \citep[see][and references therein]{desai2016,zhang2021} can spread over a large extent of the heliosphere and be observed by widely separated spacecraft. The study of these events is frequently achieved by using multiple spacecraft measurements such as (1) in the late 1970s and early 1980s using data from the Helios spacecraft and near-Earth spacecraft such as the Interplanetary Monitoring Platform IMP-8 \citep[e.g.,][]{reames1996,reames1997,lario2006}, (2) in solar cycle 24 with the twin Solar TErrestrial RElations Observatory \citep[STEREO;][]{stereo} spacecraft \citep[e.g.,][]{rouillard2012,lario2013,lario2014,lario2020,dresing2014,richardson2014,xie2019,zhuang2021}, and (3) most recently with the close-to-the-Sun measurements by Parker Solar Probe \citep[PSP;][]{psp} and Solar Orbiter \citep[SolO;][]{solo} \citep[e.g.,][]{cohen2021,kollhoff2021,lario2021,kouloumvakos2022,zhuang2022}. The combination of observations from these spacecraft can be used to answer long-standing questions about the physical processes responsible for the widespread SEP events, which may include (a) cross-field diffusion transport occurring in the corona or interplanetary space \citep[e.g.,][]{zhang2009,droge2016,zhuang2022}, (b) broad particle sources associated with wide CME-driven shock fronts \citep{reames1999,lario2016,lario2017b,kouloumvakos2022}, and/or (c) complex coronal and interplanetary magnetic field structures that may impact how SEPs spread in the heliosphere \citep[e.g.,][]{klein2008,laitinen2016,palmerio2021}.

Within the scenario of shocks accelerating particles, some studies have associated the fast access of SEPs to a broad range of heliolongitudes with large-scale disturbances observed in the low corona in the form of extreme ultraviolet (EUV) waves. EUV waves are thought to be the signature left by the footprints of the CME-driven shocks propagating in the solar corona, 
even though the nature of EUV waves is still under debate \citep[e.g.,][and references therein]{liu2014,warmuth2015}. The association between EUV waves and SEPs 
(i.e., whether or not these waves are responsible for the acceleration of the observed SEPs)
was studied in the last two decades by several authors \citep[e.g.,][]{bothmer1997,posner1997,torsti1999,rouillard2012,park2013,park2015,lario2014,lario2017a,miteva2014,prise2014,zhu2018}. For example, \citet{rouillard2012} found that the release times of energetic particles near the Sun estimated by the measurements at multiple spacecraft coincide with the time when the EUV wave expands to the spacecraft magnetic footpoints on the solar surface. Later, a study with more SEP events provided supporting evidence for such an EUV wave-SEP scenario \citep{miteva2014}. \citet{park2015} further showed a close relationship between the SEP peak flux and EUV wave speed, i.e., faster EUV waves result in higher energetic proton fluxes. However, they found that the SEP events registered at multiple spacecraft with longitudinal separation were not always consistent with the connection of the EUV waves to the spacecraft footpoints. This agrees with the result by \citet{lario2014} showing that the SEP event on 2013 April 11 was observed by two spacecraft near 1~au even though the associated EUV wave was not seen to reach the estimated magnetic footpoint of one spacecraft. Instead, the connection between the shock at higher altitudes and magnetic field lines connecting to the observers may explain the timing of the particle release.

The inconsistency between the SEP release and the EUV waves in previous studies is primarily due to the fact that, for some SEP events, the EUV wave was not seen to reach the site of the estimated magnetic footpoint of a spacecraft measuring a SEP event. However, the related question of timing still does not have a clear answer: whether or not energetic particles are immediately released when the EUV wave reaches the spacecraft footpoint. If these two times (SEP release time and EUV wave reaching the spacecraft magnetic footpoint) are not consistent, it is necessary to investigate the physical causes that influence the acceleration and release of SEPs independently of the evolution of the EUV wave. In this paper, we study a SEP event on 2021 September 28 measured by spacecraft at four different locations, i.e., by spacecraft near 1~au such as STEREO-A and spacecraft orbiting the Sun-Earth Lagrangian L1 point, as well as by PSP and SolO at inner heliospheric distances. The associated EUV wave was seen to connect to the estimated magnetic footpoints of all spacecraft but, as will be detailed below, the timing of the EUV wave connection is found to be inconsistent with the SEP release times. We note that \citet{kollhoff2021} and \citet{kouloumvakos2022} have also studied a widespread SEP event on 2020 November 29 using four spacecraft when PSP and SolO were at the distances $>$0.8~au and found that the particle release was delayed compared to the time when a connection between the EUV wave and spacecraft footpoint was established. The event studied here corresponds to a different situation, which sheds new light on the EUV wave-connection question since (a) there were three spacecraft nominally connected to closeby sites on the Sun and (b) PSP and SolO were closer to the Sun (see details in Section~\ref{sec: data_ins}). The observations of the associated eruptive phenomena in multiple wavelengths (i.e., radio, white-light, and EUV) can be used to investigate the relationship between the coronal shock and the acceleration and release of SEPs.

The rest of the paper is structured as follows. In Section~\ref{sec: data_ins}, we describe the data and instruments used in this study. Section~\ref{sec: obs_res} introduces the observations and analysis of the SEP event and the associated EUV wave and coronal shock. In Section~\ref{sec: dis}, we discuss the acceleration of SEPs associated with the coronal shock and the uncertainties in the timing comparison. Section~\ref{sec: sum} summarizes the main results of the present study.

\section{Data and Instrumentation}\label{sec: data_ins}

\subsection{Instrumentation}
In this study, we investigate the SEP event and its associated eruptive phenomena using both remote-sensing instruments at multiple wavelengths and in-situ instruments from multiple locations. In particular, to study the eruptive phenomena from radio to EUV wavelengths passing by white light, we use (a) the Radio and Plasma Wave \citep[WAVES;][]{waves} experiment on board the Wind spacecraft and radio data from CALLISTO's ASSA station in Australia \citep{ecallisto}, (b) the Large Angle and Spectrometric Coronagraph on board the SOlar and Heliospheric Observatory \citep[SOHO/LASCO;][]{soho} and the coronagraphs COR1 and COR2 \citep{secchi} on board STEREO Ahead (hereafter STA), and (c) the Extreme Ultraviolet Imager \citep[EUVI;][]{euvi} on board STA and the Atmospheric Imaging Assembly \citep[AIA;][]{aia} on board the Solar Dynamic Observatory (SDO).

To investigate the SEP events in situ from different locations, we use (a) the Electron and Proton Telescope (EPT) and the High-Energy Telescope (HET) of the Energetic Particle Detector \citep[EPD;][]{epd} on board SolO, (b) one of the Low-Energy Telescopes (LET1) and the High-Energy Telescope (HET) of the Integrated Science Investigation of the Sun (IS$\odot$IS) instrument suite \citep{isois} (version-13 data) on board PSP, (c) LET \citep{let} and HET \citep{het} on board STA, and (d) the Energetic and Relativistic Nuclei and Electron Experiment \citep[ERNE;][]{erne} on board SOHO and the Three-Dimensional Plasma and Energetic Particle Investigation suite of instruments on board Wind \citep[3DP;][]{3dp} at L1. SolO/EPT and SolO/HET each have four apertures (termed Sun, ASun, North, and South) that scan different sky regions \citep[see details in][]{epd}. Both PSP/LET1 and PSP/HET are double-ended telescopes. During this SEP event, the primary axis of PSP was consistently pointed at the Sun. However, for momentum management and communication purposes, the spacecraft performed several 180$^\circ$ rotations about the primary axis. As a result, the two LET1 ends (LET-A and LET-B) were nominally pointed $45^\circ$ west (or east due to the rotation) and $135^\circ$ east (or west due to the rotation) of the Sun–spacecraft line, respectively. Similarly, the PSP/HET two ends (HET-A and HET-B) were nominally pointed $20^\circ$ west (east) and $160^\circ$ east (west) of the Sun–spacecraft line. No substantial effect of these rotations was visible in the measured energetic particle intensities. STA/LET has 16 sectors, divided into two major ends having the center axis pointed $45^\circ$ east of and $135^\circ$ west of the Sun-spacecraft line, respectively. Wind/3DP provides particle fluxes binned at eight different pitch angles. We note that SolO/EPT and Wind/3DP measure ions without distinguishing their species, and presumably the measured fluxes are dominated by the most abundant protons. Thus, we also use the term proton to describe such measurements hereafter.

Interplanetary magnetic field (IMF) and solar wind plasma measurements were obtained from (a) the magnetometer \citep[MAG;][]{solo_mag} and the Solar Wind Analyser \citep[SWA;][]{solo_swa} on board SolO, (b) the Electromagnetic Fields Investigation \citep[FIELDS;][]{fields} on board PSP, (c) the magnetometer \citep{impact} and the Plasma and Suprathermal Ion Composition (PLASTIC) Investigation \citep{plastic} on board STA, and (d) the solar wind plasma \citep[SWE;][]{wind_swe} and magnetic field data \citep[MFI;][]{wind_mfi} on board Wind.

\subsection{Spacecraft Location}\label{sec: sc_loc}
\begin{figure}[!hbt]
    \centering
    \includegraphics[width=1.\textwidth]{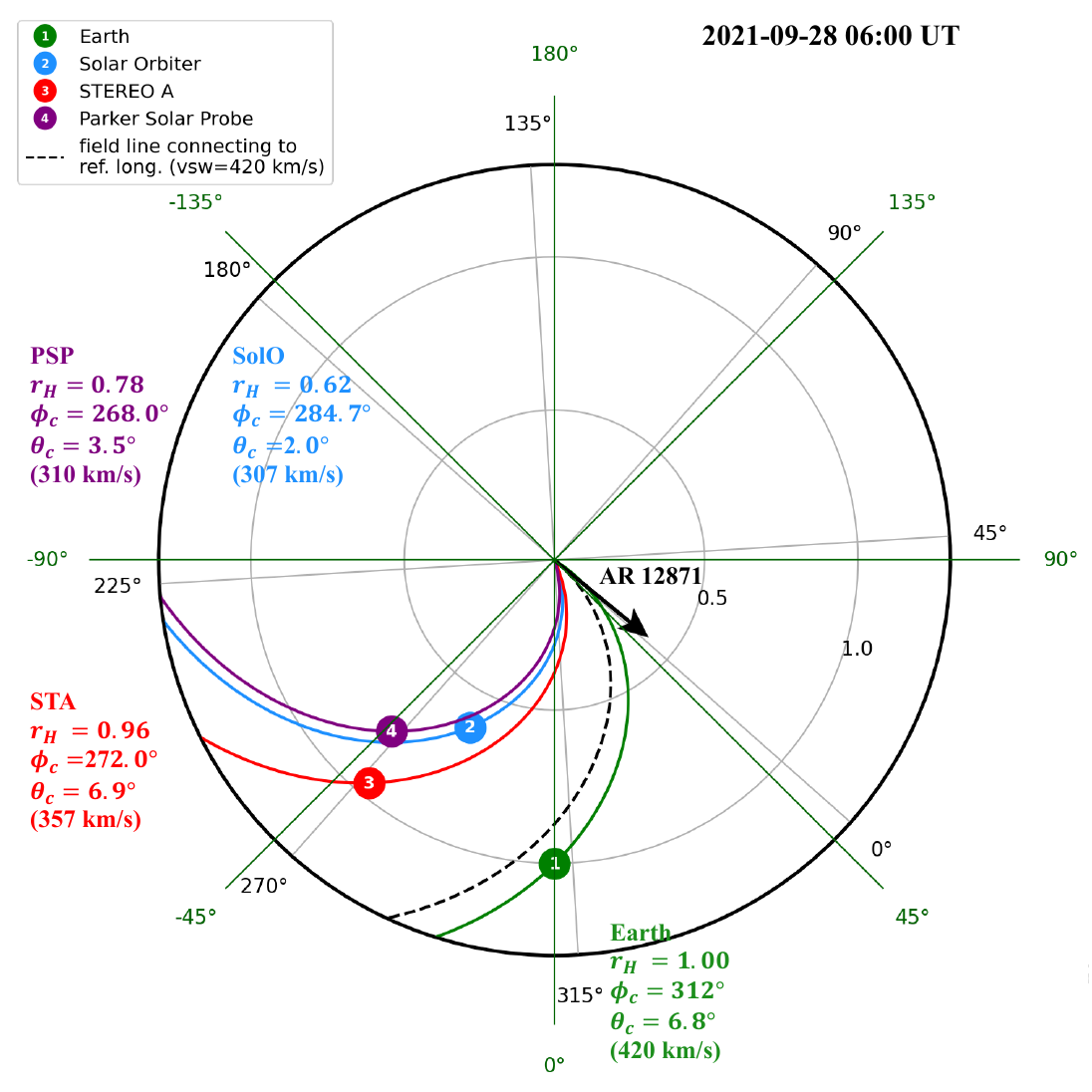}
    \caption{Locations of SolO, PSP, STA, and Earth at 06:00~UT on 2021 September 28 in both Carrington (black) and Stonyhust (green) coordinates together with nominal Parker spiral IMF lines connecting each spacecraft with the Sun. The black arrow indicates the longitude of AR~12871 and its associated nominal spiral IMF line is shown by the dashed black curve. The spacecraft and Earth coordinates (heliocentric distance $r_H$, Carrington latitude $\theta_C$ and longitude $\phi_C$) and the $v_{\rm sw}$ values (in brackets) used to compute the nominal Parker spiral IMF lines are also listed.
    }
    \label{fig: sc_loc}
\end{figure}
The SEP event on 2021 September 28 was observed at the four locations where SolO, PSP, STA, and L1 were located (we use L1 to indicate the location of the SOHO and Wind spacecraft). During this time, SolO and PSP were at heliocentric distances of 0.62 and 0.78~au, respectively, and the largest longitudinal separation between any two spacecraft was $44^\circ$. Figure~\ref{fig: sc_loc} shows the spacecraft locations at 06:00 UT on 2021 September 28 in the ecliptic plane in both Carrington (black) and Stonyhurst (green) coordinates generated using the Solar-Mach tool \citep[][\url{https://solar-mach.github.io/}]{solar-mach}. The solid curves are the nominal Parker spiral IMF lines connecting the spacecraft and Sun. The nominal Parker spiral of a spacecraft is estimated using a constant solar wind speed ($v_{\rm sw}$) which is an average of the in-situ plasma measurements within a six-hour duration before the SEP onset time. Note that there were no plasma measurements with good-quality flags by the Solar Wind Electrons Alphas and Protons \citep[SWEAP;][]{sweap} on board PSP during this time, and thus we assume a $v_{\rm sw}$ of 310~km\,s$^{-1}$. This assumption is based on the solar wind measured at the nearby spacecraft: $v_{\rm sw}$ was around 300~km\,s$^{-1}$ at SolO and 350~km\,s$^{-1}$ at STA during the day of September 28. The spacecraft coordinates (heliocentric distance $r_H$, Carrington latitude $\theta_C$ and longitude $\phi_C$), the $v_{\rm sw}$ values used to compute the nominal Parker magnetic field lines, and nominal field line lengths ($l$) connecting each spacecraft with the Sun are listed in Table \ref{tab: sep_info}. The in-situ magnetic field and solar wind plasma measurements are also shown in Figure~\ref{fig: in-situ-para} in the Appendix.

The longitude of the nominal magnetic footpoint ($\phi_F$) of each spacecraft is estimated assuming nominal Parker spiral field lines connecting each spacecraft with the Sun (the values of the footpoint longitudes in Carrington coordinates are provided in Table \ref{tab: sep_info}). It shows that the nominal magnetic connections of SolO, PSP, and STA were very close in longitude (e.g., the longitudinal separation between the footpoints of PSP and STA was $7^\circ$). In contrast, the nominal footpoint of the L1 spacecraft was separated from the region where SolO, PSP, and STA established magnetic connections by around $40^\circ$ in longitude. Such a distribution of multiple spacecraft is useful for studying not only the longitudinal extent of the SEP event but also the particle transport along closeby IMF lines from the source region at different heliocentric distances.

\section{Observations and Results}\label{sec: obs_res}
The solar origin of the SEP event was associated with a CME eruption and a C1.6-class flare from active region 12871 (location: S27W51 in Stonyhurst coordinates). The onset time of the C1.6 flare was at 05:54~UT on 2021 September 28. The CME was first seen by LASCO/C2 at 06:24~UT on September 28 already at a height of 2.6~$R_\odot$. This CME propagated with a plane-of-sky speed of about 600~km~s$^{-1}$ (as reported in the Coordinated Data Analysis Workshop (CDAW) Data Center; \url{https://cdaw.gsfc.nasa.gov/CME_list/index.html}) and drove a shock with a final speed of about 900~km~s$^{-1}$ (see Section \ref{sec: shock}).

\subsection{SEP In-Situ Observations}\label{sec: in-situ-sep}
\begin{figure}[!hbt]
    \centering
    \includegraphics[width=0.9\textwidth]{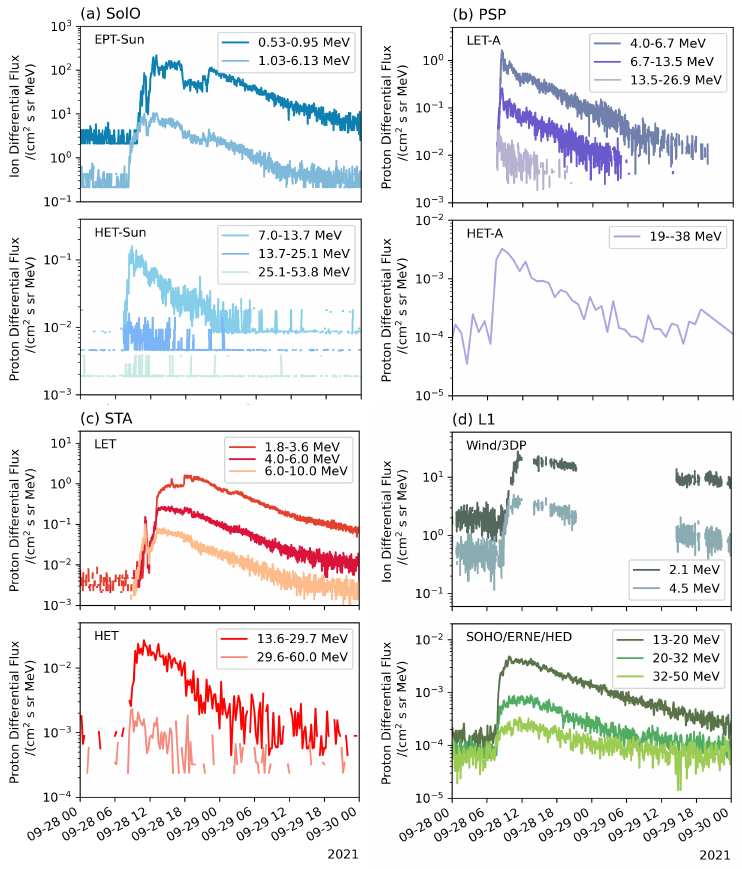}
    \caption{Proton and ion intensities at different energies of the SEP event at (a) SolO, (b) PSP, (c) STA, (d) L1 (i.e., at SOHO and Wind). Except for the 5-minute averaged data of STA/HET and the 15-minute averaged data of PSP/HET, all instruments have 1-minute averaged data points. At SolO, the sunward telescopes of EPT and HET are used. At PSP, the A-side telescopes of LET and HET are used. At STA/LET and Wind/3DP, the data are sector-averaged.
    }
    \label{fig: sep_obs}
\end{figure}
Figure~\ref{fig: sep_obs} shows ion and proton intensities at different energies collected by, (a) the sunward apertures of EPT and HET telescopes on SolO, (b) LET-A and HET-A on PSP, (c) the sector-averaged intensities measured by LET and HET on STA, and (d) the sector-averaged intensities collected by Wind/3DP and the intensities collected by SOHO/ERNE. The highest energies at which proton intensity enhancements were observed were about $\sim$50~MeV. 

We focus on the SEP onset phase in this study. In order to estimate the particle release time, we use a time-shifted analysis (TSA) which assumes scatter-free particles transporting along nominal Parker spiral IMF lines. We only focus on the proton measurements at relatively high energies because the scattering effects on high-energy protons are assumed to be less important than on low-energy protons. The particle release time in TSA is $t_{\rm SPR}(E) = t_{\rm onset}(E) - 8.33\frac{\rm min}{\rm au}l/\beta(E)$, where $t_{\rm onset}(E)$ is the start time of the particle flux enhancement at kinetic energy $E$, $1/\beta(E)$ is the reciprocal of the speed of particles with kinetic energy $E$, and $l$ is the length of the nominal Parker spiral line. The values of $l$ of the four locations are listed in Table \ref{tab: sep_info}. 

To estimate $t_{\rm onset}$, we use three methods: (1) the so-called 3-$\sigma$ method \citep{krucker1999}, (2) the Poisson-CUSUM method \citep{poisson-cusum}, and (3) visual identification. In general, the first method relies on the rise of the particle intensity relative to the pre-event background intensity by a certain threshold. The second method assumes that the pre-event background is relatively steady and accumulates the difference between the measurements and a reference value related to the background intensity. The SEP onset time is then determined by the time when the accumulation of the difference exceeds a certain threshold, and such exceedance is required to remain for the following (e.g., 30) data points after the first exceedance occurs. Compared to the first method, CUSUM avoids some false detections of the SEP onset caused by the disturbance of the background intensity. However, these two methods require high-quality pre-event background intensity data. If this criterion cannot be met (see e.g., proton intensities at PSP/LET-A as shown in panel~(b) of Figure~\ref{fig: sep_obs}), then the third method is used. Overall, as for the proton fluxes with pre-event measurements, the three methods are used simultaneously (e.g., SOHO/ERNE), and the average and standard deviation values represent the corresponding onset time and uncertainty; otherwise, only the third method is used with a presumed uncertainty of 3 minutes (triple the temporal resolution of the data, e.g., for PSP/LET). 

The identified onset times at different energies and spacecraft are listed in Table \ref{tab: sep_info}. Using the geometric mean of a specific energy range, $t_{\rm SPR}$ is subsequently estimated to be 06:52~UT for 7--13.7~MeV protons at SolO, 06:44~UT for 6.7--13.5~MeV protons at PSP, 07:16~UT for 6--10~MeV at STA, and 06:55~UT for 13--20~MeV protons at L1, respectively, with uncertainties inherited from the $t_{\rm onset}$ uncertainties (Table \ref{tab: sep_info}). When comparing the particle release time with the EUV wave and the shock observations in remote-sensing images obtained from observers near 1~au, the light-travel time to 1~au of $\sim$8.3 minutes is added to $t_{\rm SPR}$. We note that $t_{\rm SPR}$ at STA was delayed compared to those at SolO and PSP, which may be due to the fact that STA did not register the first-arrival proton (see Section~\ref{sec: dis_uncer}). The path length effect on $t_{\rm SPR}$ is also discussed in Section~\ref{sec: dis_uncer}.

We have also used the velocity-dispersion analysis \citep[VDA;][]{vainio2013} to estimate the particle release time and path length. For this, the measurements at relatively lower energies (but still $>$0.5~MeV) are incorporated to ensure enough data points are incorporated for the estimation (due to the low signal level at higher energies). The estimated release times in VDA were quite similar to the TSA results, and the path lengths at SolO and PSP were close to the nominal Parker spiral lengths (not shown here).

\subsection{EUV Wave Observations}\label{sec: euv}
\begin{figure}[!hbt]
    \centering
    \includegraphics[width=\textwidth]{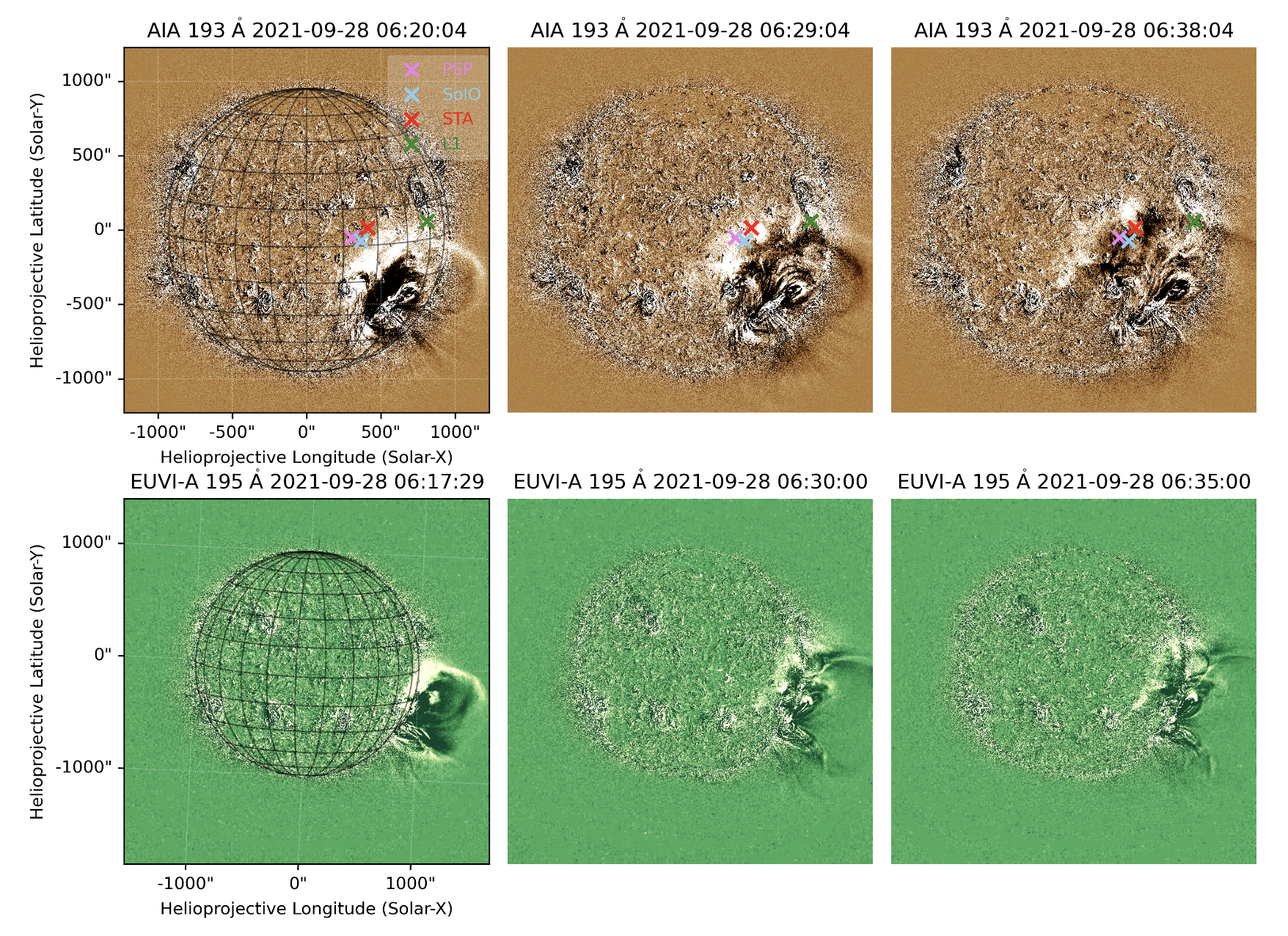}
    \caption{Running difference images of the EUV wave observed by (top panels) SDO/AIA in 193~\AA~and (bottom panels) STA/EUVI in 195~\AA~at similar times. The four crosses in the top panels indicate the nominal magnetic footpoint of the four spacecraft.
    }
    \label{fig: euv}
\end{figure}

Figure \ref{fig: euv} shows the EUV wave evolution/expansion 
observed by SDO/AIA in 193~\AA~in the top panels and STA/EUVI in 195~\AA~in the bottom panels. Nominal magnetic footpoints of the four spacecraft are marked in the AIA images shown in the top panels of Figure~\ref{fig: euv}. In the AIA field-of-view (FOV), the major part of the EUV wave propagated mostly in the northeast direction, while its southward propagation was not as obvious. We note that there were two fronts of the EUV wave, in which the outside (northern) one was faster with a speed of around 315~km~s$^{-1}$ and the inner (southern) one was slower with a speed of around 110~km~s$^{-1}$. The ratio of the inner front speed over the outer front speed is found to be around $1/3$. This leads us to conclude that the outer front was related to the EUV wave front, and the inner one to the coronal magnetic field and material expelled by the CME expansion, as proposed by \citet{chen2002} and \citet{downs2012}. We then record the time when the EUV wave intercepted the spacecraft nominal magnetic footpoints (listed in Table~\ref{tab: sep_info}). The uncertainties of using nominal footpoint in estimating the footpoint-connecting time are discussed in Section \ref{sec: dis}. The outer front of the EUV wave (interpreted as a shock front) was seen to be nearly the west limb of the Sun as seen from STA. This means that the plane-of-sky of STA can provide a good visualization of the structure and also helps to perform the shock reconstruction (see next section).

\subsection{Shock Model}\label{sec: shock}
The existence of the CME-driven shock is confirmed by the observations of the EUV wave and type II radio burst (as shown in Figure~\ref{fig: radio} below). 
\begin{figure}[!hbt]
    \centering
    \includegraphics[width=0.9\textwidth]{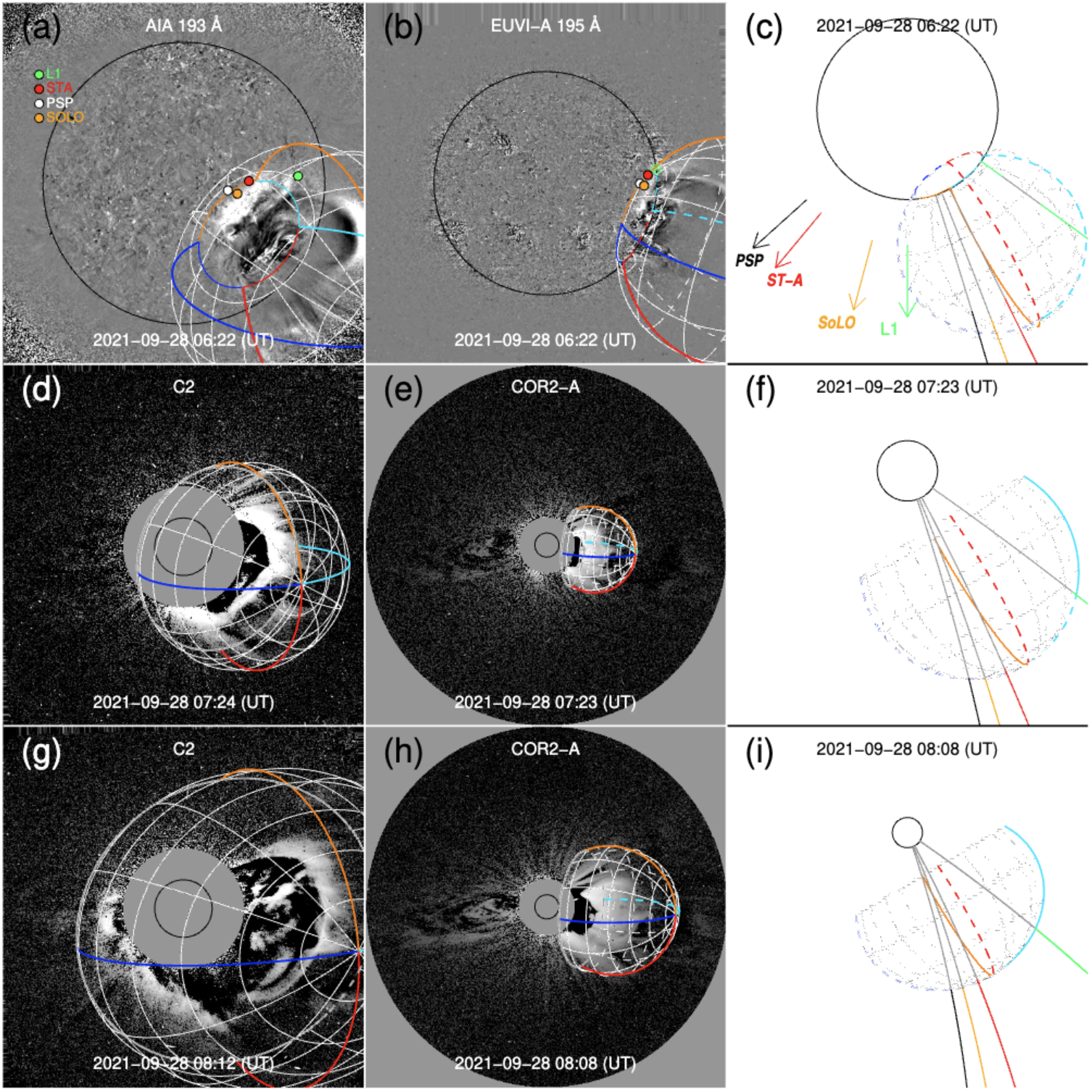}
    \caption{Evolution of the 3D geometry of the coronal shock wave. The left and middle columns show the observations taken from (a) SDO/AIA~193~\AA, (b) STA/EUVI~195~\AA, (d, g) SOHO/LASCO/C2, and (e, h) STA/COR2. The 3D reconstructions of the shock are overlapped on these images. The footpoints of the nominal magnetic field lines of the four spacecraft, near-Earth observer (L1), SolO, STA, and PSP, on the solar surface, are shown with filled circles in panels (a) and (b). The right columns show the ecliptic plane seen from the north to illustrate the shock propagation with respect to the nominal magnetic field lines connecting the Sun and the four spacecraft. The directions toward each spacecraft are represented with arrows in panel (c). The open circles in the panels of the right column refer to the Sun, and the FOVs of these panels from top to bottom are 2\,$R_{\odot}$, 6\,$R_{\odot}$, 12\,$R_{\odot}$, respectively.
    }
    \label{fig: shock}
\end{figure}
In order to determine the shock geometry and kinematics in the corona, we use a shock geometrical ellipsoid model \citep{kwon2014} that combines SOHO/LASCO, SDO/AIA, STA/EUVI, and STA/COR observations from multiple viewpoints to fit the outermost front of the CME interpreted as the shock wave driven by the CME \citep{ontiveros2009}. This shock model uses seven free parameters, including the latitude, longitude, and height of the ellipsoid center, the lengths of the three semi-principle axes, and the ellipsoid rotation angle. One can refer to \citet{kwon2014} for more details.

Figure~\ref{fig: shock} shows the observations of, from top to bottom, (1) the EUV wave in (a) SDO/AIA and (b) STA/EUVI, (2) the white-light CME in (d) SOHO/LASCO/C2 and (e) STA/COR2, and (3) the white-light CME in (g) LASCO/C2 and (h) STA/COR2 at three different time steps and overlapped with the reconstructed shock ellipsoid structure. We note that the shock signature in white-light coronagraph images is not very distinct in all directions, and thus the observations in EUV passband have become more significant in the reconstruction of the shock geometry (e.g., the footprint of the shock on the solar surface has been matched with the EUV wave). Nominal magnetic field lines connecting the shock front and the four spacecraft are shown in panels (c), (f), and (i) (viewing from the north ecliptic pole). Based on this model, the shock nose propagated along the latitude of S20 and the longitude of W60 in the Heliocentric Earth Equatorial coordinates with a final speed of around 900~km~s$^{-1}$.

{We then obtain the height of the shock front at the point of the shock front that intercepted the field line connecting to each spacecraft \citep[also known as the Connecting-with-the-Observer-Point, or cobpoint, after][]{heras1995}. The results are shown in panel~(b) of Figure~\ref{fig: time}. The velocity of the shock in three-dimensional (3D) space was derived by considering the minimum distance trajectory between two ellipsoid structures modeled at two different time stamps as done in \citet{kwon2017}. Then, the speed of the shock at the cobpoints of the four spacecraft was obtained from the 3D velocity field. The speeds at the cobpoints are shown in Figure~\ref{fig: time}(c) and those of SolO, PSP, and STA were quite similar, while the speed for L1 was lower because the related cobpoint was quite distant from the shock nose (see panel~(i) in Figure~\ref{fig: shock}). Furthermore, based on the reconstructed shock geometry and nominal IMF lines, we can estimate the angle between the magnetic field and shock normal ($\theta_{Bn}$) at each cobpoint of the observer, ranging between around $10^\circ$ and $45^\circ$ during the particle release times estimated from the four-observer measurements.

\subsection{Time Sequence of the Events}\label{sec: time}
We compare the particle release time (verticle bars) and the time when the EUV wave connected to the spacecraft footpoint (filled squares) in panels~(b) and (c) of Figure~\ref{fig: time}.
\begin{figure}[!hbt]
    \centering
    \includegraphics[width=0.95\textwidth]{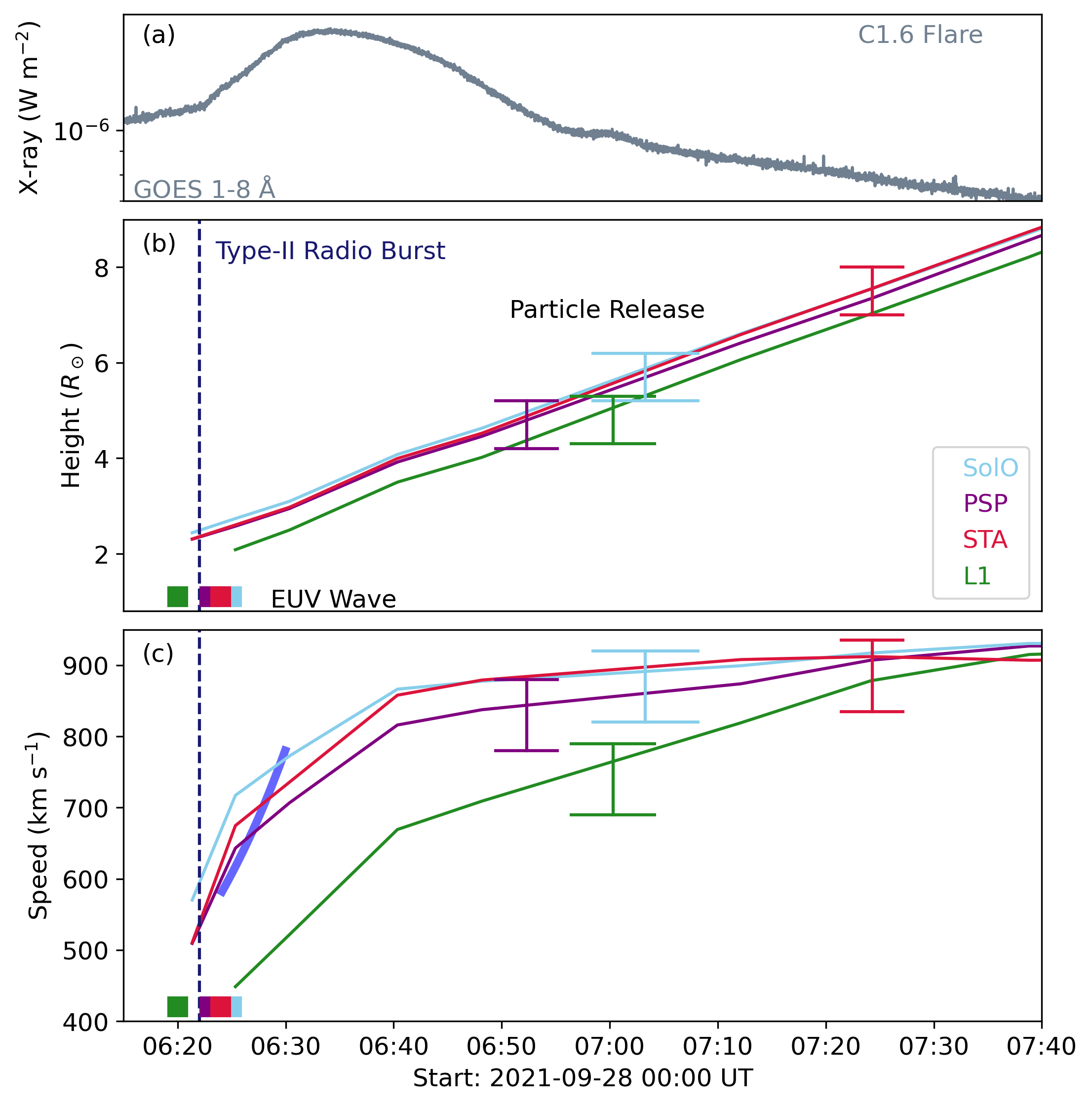}
    \caption{(a): GOES soft X-ray flux. (b) and (c): height and speed vs. time of the shock at the SolO (blue), PSP (purple), STA (red), and L1 (green) cobpoints, respectively. The time when the EUV wave connected to the nominal magnetic footpoint of the four spacecraft (filled squares), the type-II radio burst onset time (vertical dashed line), and the particle release time with 8.3 minutes added (solid lines with error bars) are shown in panels~(b) and (c). The blue thick line in panel~(c) indicates the shock speed derived from the radio observations (Section~\ref{sec: dis_radio}).
    }
    \label{fig: time}
\end{figure}
We find that the estimated particle release times were delayed by around 30 to 60 minutes compared to the EUV wave connection times for all four spacecraft. Such a delay is significant even when considering the uncertainties of both the EUV wave connection times and the particle release times (see Section~\ref{sec: dis_uncer}). We note that the particle release time at STA displayed the largest delay, and this may be related to the fact that the particle instruments on STA did not register the first particles arriving at STA location (Section \ref{sec: dis}). At the particle release times, the cobpoint heights were around 5--6~$R_\odot$ for SolO, PSP, and L1, which is consistent with previous results of the acceleration and release of particles occurring when the shock propagates to around 4~$R_\odot$ \citep[e.g.,][]{gopalswamy2010,li2012}. The cobpoint height for STA at the particle release time estimated was around 7.5~$R_\odot$, but this estimation may be influenced by the viewing angles of the STA/LET sectors (see Section~\ref{sec: dis_uncer}).

We turn our attention to panel~(c) of Figure~\ref{fig: time}. The shock was not extremely fast in the low corona, since its speed was around 500~km\,s$^{-1}$ during the EUV wave expansion time (at the L1 cobpoint, the speed was slightly lower). The shock then accelerated to reach a speed of 800--900~km~\,s$^{-1}$ at the particle release times. Panel~(a) of Figure~\ref{fig: time} shows the time series of the GOES soft X-ray flux in 1--8~\AA. The variation of the X-ray flux profile is consistent with the variation of the shock speed profile. Since this shock was driven by the associated CME, such a relationship is reasonable due to the close coupling between flares and CMEs \citep[e.g.,][]{zhang2004,gou2020,zhuang2022b}. 

\section{Discussion}\label{sec: dis}
In this section, we first introduce the radio burst observations and the model used to estimate the coronal and shock properties from the type II radio burst. We then discuss the particle acceleration associated with the evolution of the shock which is thought to be the main reason for the delayed particle release time. We finally discuss the uncertainties in estimating the times.

\subsection{Radio Observations}\label{sec: dis_radio}
A type II solar radio burst with the onset time at around 06:23~UT was observed by CALLISTO/ASSA as shown in panel~(a) of Figure~\ref{fig: radio}. 
\begin{figure}[!hbt]
    \centering
    \includegraphics[width=\textwidth]{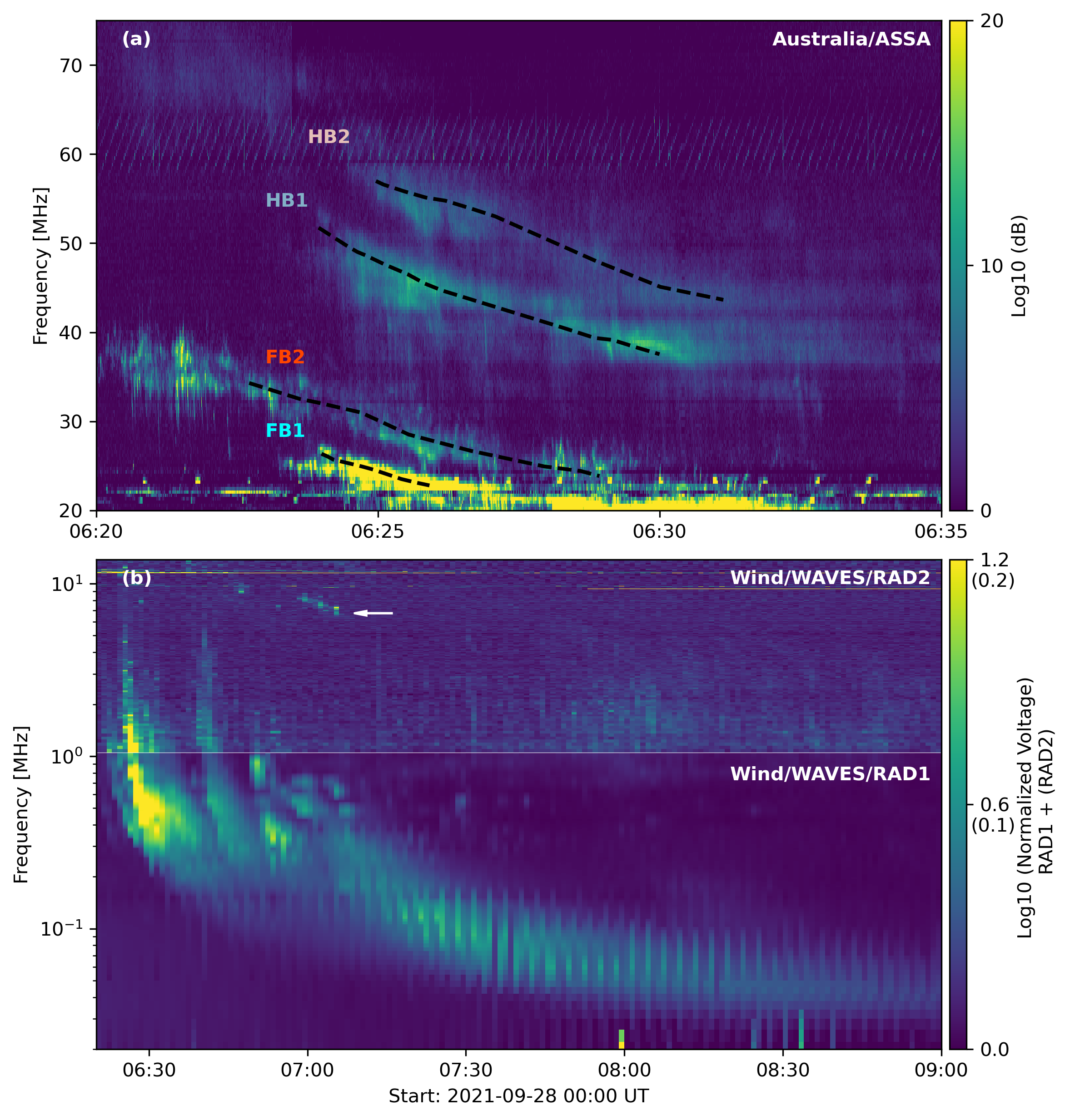}
    \caption{Radio dynamic spectra (background-subtracted) of the type II and type III radio bursts observed by (a) CALLISTO/ASSA and (b) Wind/WAVES. The black dashed tracks in panel~(a) outline the data points selected to estimate the drift of each type II lane. In panel~(b), the arrow indicates the type II radio burst observed by Wind/WAVES, and different scale values of the color-bar are used for RAD1 and RAD2 (in brackets).
    }
    \label{fig: radio}
\end{figure}
Fundamental and harmonic bands (FB and HB, respectively) can be seen in the type II burst. An intriguing phenomenon here is that the two bands were split in two lanes (FB1 with FB2, and HB1 with HB2), known as band splitting \citep{roberts1959}. The mechanism of this band splitting is still under debate \citep[see discussions in e.g.][]{chrysaphi2018}.  Two prominent theories involve two opposing arguments: (1) the emission from the upstream (ahead) and downstream (behind) regions of the shock front \citep{smerd1975}, or (2) the emission from different parts upstream of the shock front \citep{1983ApJ...267..837H}.
The dashed curves outline the frequency drift of the four lanes, consisting of data points taken every 10 to 60 seconds based on different signal strengths. Before the drifting type II emissions, there were stationary emissions between 06:20~UT and 06:23~UT, which could be a pre-cursor or part of a transitioning type II burst \citep[e.g.,][]{chrysaphi2020}. 

Panel~(b) of Figure~\ref{fig: radio} shows the radio spectrum recorded by Wind/WAVES (where the time differs from that of the CALLISTO/ASSA observation). The type II burst is barely seen by WIND at about 10~MHz, marked by the white arrow. Type III radio bursts are visible, indicating the release of electrons associated with the solar eruption. We note that during this SEP event, there were no radio imaging observations that enable us to estimate the source regions of the radio burst. Nevertheless, we can assume that the upstream-downstream model \citep{smerd1975} has led to the frequency split between the type II lanes ($\Delta f/f=\frac{f_u-f_l}{f_l}$, where $f_u$ and $f_l$ are the upstream and downstream frequencies of the fundamental lane).  This assumption allows us to obtain an estimate of the coronal and shock conditions in the low corona. One can refer to Appendix~\ref{sec: app_radio} for detailed descriptions of how we have estimated the upstream shock Alfv\'{e}nic Mach number, shock speed, and upstream magnetic field strength.

\begin{figure}[!hbt]
    \centering
    \includegraphics[width=0.96\textwidth]{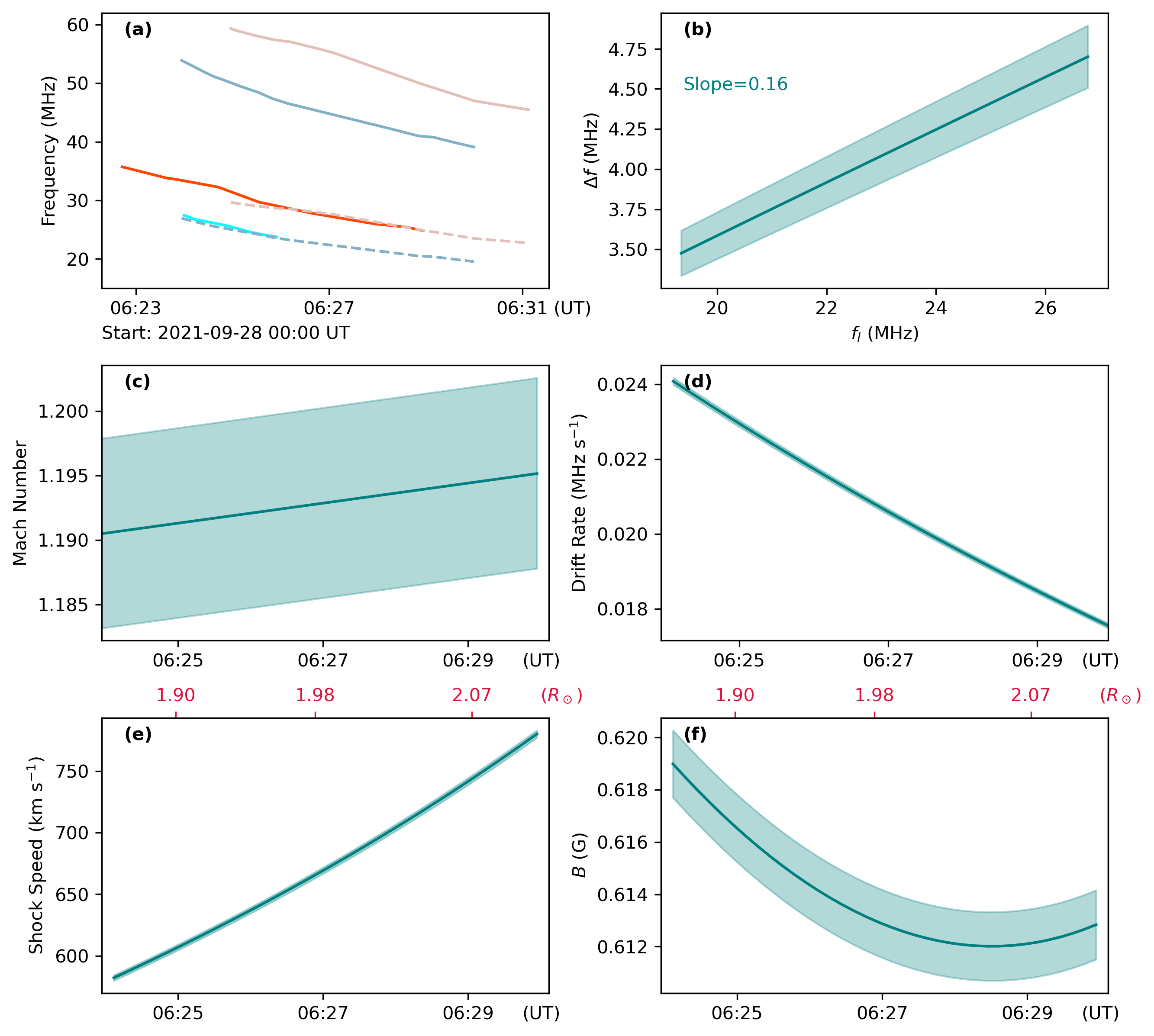}
    \caption{The properties of the shock and coronal magnetic field derived from the radio observations. (a): the four radio lanes taken from the four dashed tracks in Figure~\ref{fig: radio} (solid) and the half frequencies of the two harmonic lanes (dashed). (b): $\Delta f$ versus $f_l$. (c): shock Alfv\'{e}nic Mach number versus time. (d): approximated drift rate of FB1 lane versus time.
    (e): shock speed versus time with the associated shock front height (obtained using the one-fold Newkirk model) listed at the top. (f): upstream magnetic field strength versus time with the shock front height. The shaded regions in panels~(b)--(f) indicate the associated uncertainties. More details can be found in the main text.
    }
    \label{fig: radio_model}
\end{figure}
Figure~\ref{fig: radio_model} shows the results of this model. The four dashed tracks as shown in Figure~\ref{fig: radio} are re-plotted by the solid curves in panel~(a). The dashed curves indicate the half frequencies of the harmonic bands, which are consistent with the fundamental bands. In the following calculation, we do not directly use the selected data points because $f_l$ of the FB1 lane and $f_u$ of the FB2 lane cannot always be selected at the same time step. Instead, we use a linear fit of the logarithm of frequency versus time to represent the associated frequency drift of each lane. Panel~(b) shows $\Delta f$ versus $f_l$ based on the linear fit, in which the shaded region indicates the uncertainties coming from the fit (also for the remaining panels). The slope of $\Delta f/f_l=0.16$ is consistent with previous statistical studies as shown in, e.g., \citet{vrsnak2001}. Such a $\Delta f/f_l$ leads to an estimate of the shock compression ratio $X$ to be $\sim$1.38. Combining the band-splitting properties and the Rankine-Hugoniot conditions, the upstream shock Alfv\'{e}nic Mach number is derived and shown in panel~(c), which was around 1.19 during the type II burst. This highlights the fact that the Mach number was already greater than 1 at the beginning of the radio burst. However, in the next section, we illustrate that, during the type II emissions, the shock may not have been strong enough to be responsible for the efficient acceleration and following release of energetic particles as the particle release times are found to be delayed compared to the onset time of the type II radio burst as shown in Figure~\ref{fig: time}.

The frequency drift rate ($df_l/dt$) of the FB1 lane is shown in panel~(d) of Figure~\ref{fig: radio_model}. Combining the drift rate and the one-fold Newkirk coronal model \citep[see Appendix~\ref{sec: app_radio};][]{newkirk1961}, the shock speed is calculated in panel~(e) of Figure~\ref{fig: radio_model}, increasing from around 580~km\,s$^{-1}$ at 06:24~UT to 780~km\,s$^{-1}$ at 06:30~UT. The associated shock front height estimated using the one-fold Newkirk model is marked at the top axis (in red). We note again, without radio imaging information, we cannot pinpoint the exact location of the radio emission source, e.g., whether it was the shock nose, the shock flanks, or any other neighboring region. However, the radio emission information is still useful: the blue thick bar in panel~(c) of Figure~\ref{fig: time} indicating the shock speed derived from the radio observations is generally consistent with the results derived independently from the shock model. Finally, by combining the estimated shock speed and Mach number, the magnetic field strength in the corona is estimated to be around 0.61~Gauss at the height of $\sim$2~$R_\odot$ as shown in panel~(f) of Figure~\ref{fig: radio_model}.

\subsection{Particle Acceleration with the Evolution of the Shock}\label{sec: dis_acc}
The delayed particle release with respect to the estimated EUV wave arrival time to the spacecraft magnetic footpoints may be caused by multiple factors, including (1) the shock acceleration is not efficient initially, (2) there is a delay between the particles being accelerated and finally released, and/or (3) the IMF and other transients influence the particle propagation (discussed in Section~\ref{sec: dis_uncer}). 
Combining a shock model reproduction from remote-sensing observations and magnetohydrodynamic (MHD) simulations of the coronal background, \citet{lario2017a} found that the SEP release time does not always coincide with time when the estimated fast magnetosonic Mach number first exceeds a given threshold. \citet{lario2017a} discussed the effects that the uncertainties when estimating the large-scale structure of the shock, the ambient coronal background, the shock parameters, and the particle release times have in establishing the association between the shock evolution and the acceleration of SEPs. In this paper, we reevaluate the possible link between the shock strength (estimated through the Alfv\'{e}nic Mach number as the main factor responsible for the efficient particle acceleration) and their subsequent release into interplanetary space. The combination of remote-sensing and in-situ observations during the 2021 September 28 event allows us to carefully examine this association.

\begin{figure}[!hbt]
    \centering
    \includegraphics[width=\textwidth]{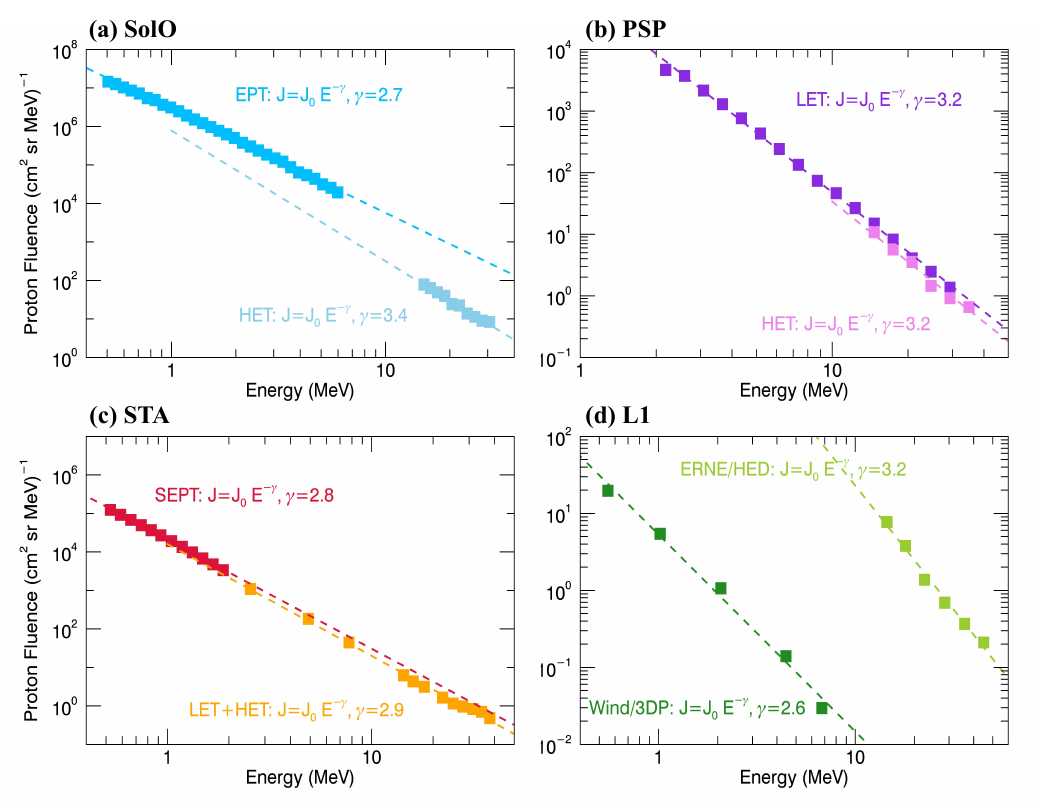}
    \caption{Event-integrated energy spectrum at (a) SolO, (b) PSP, (c) STA, and at (d) L1 (using SOHO and Wind). The dashed lines indicate the exponential fits for different instruments by different colors. More details can be found in the main text.
    }
    \label{fig: spectrum}
\end{figure}
We first look at the event-integrated energy spectrum which helps understand the SEP properties. Figure~\ref{fig: spectrum} shows the event-integrated energy spectrum for the four spacecraft, in which the proton flux in each energy range is corrected by a background level which is the average over the six hours before the SEP onset. The integration time used to compute the energetic particle spectra is different for the four spacecraft (covering the whole enhancement duration): between 06:00~UT on September 28 and 12:00~UT on September 29 for SolO, between 07:00~UT on September 28 and 12:00~UT on September 29 for PSP, between 08:40~UT on September 28 and 12:00~UT on October 1 for STA, and between 08:00~UT on September 28 and 00:00~UT on September 30 for L1. The instruments used to make this plot are the same as those used in Figure~\ref{fig: sep_obs} with the addition of the sunward aperture of the Solar Electron and Proton Telescope \citep[SEPT;][]{sept} on board STA. We use an exponential fit ($J=J_0 E^{-\gamma}$, where $J$ is the fluence) to derive the spectra index ($\gamma$). The study of the inter-calibration between the particle intensities measured by different instruments on each spacecraft is beyond the scope of the present study; therefore we separately fit the fluence-energy data points for the instruments in lower and higher energy ranges at each spacecraft (indicated by the dashed lines in different colors). We find that the spectrum in the lower energy range was slightly harder than that in the higher energy range for SolO and L1 but remains the same for PSP and STA. We focus on protons at higher energies and find that as for the four spacecraft, $\gamma$ was around 3. We note these $\gamma$ values for the high-energy detectors derived based on the fluences integrated over the entire SEP event were tested to be similar to the values obtained by integrating only the first four or six hours, even though the derived shock parameters changed quickly within the first few hours.

\begin{figure}[!hbt]
    \centering
    \includegraphics[width=\textwidth]{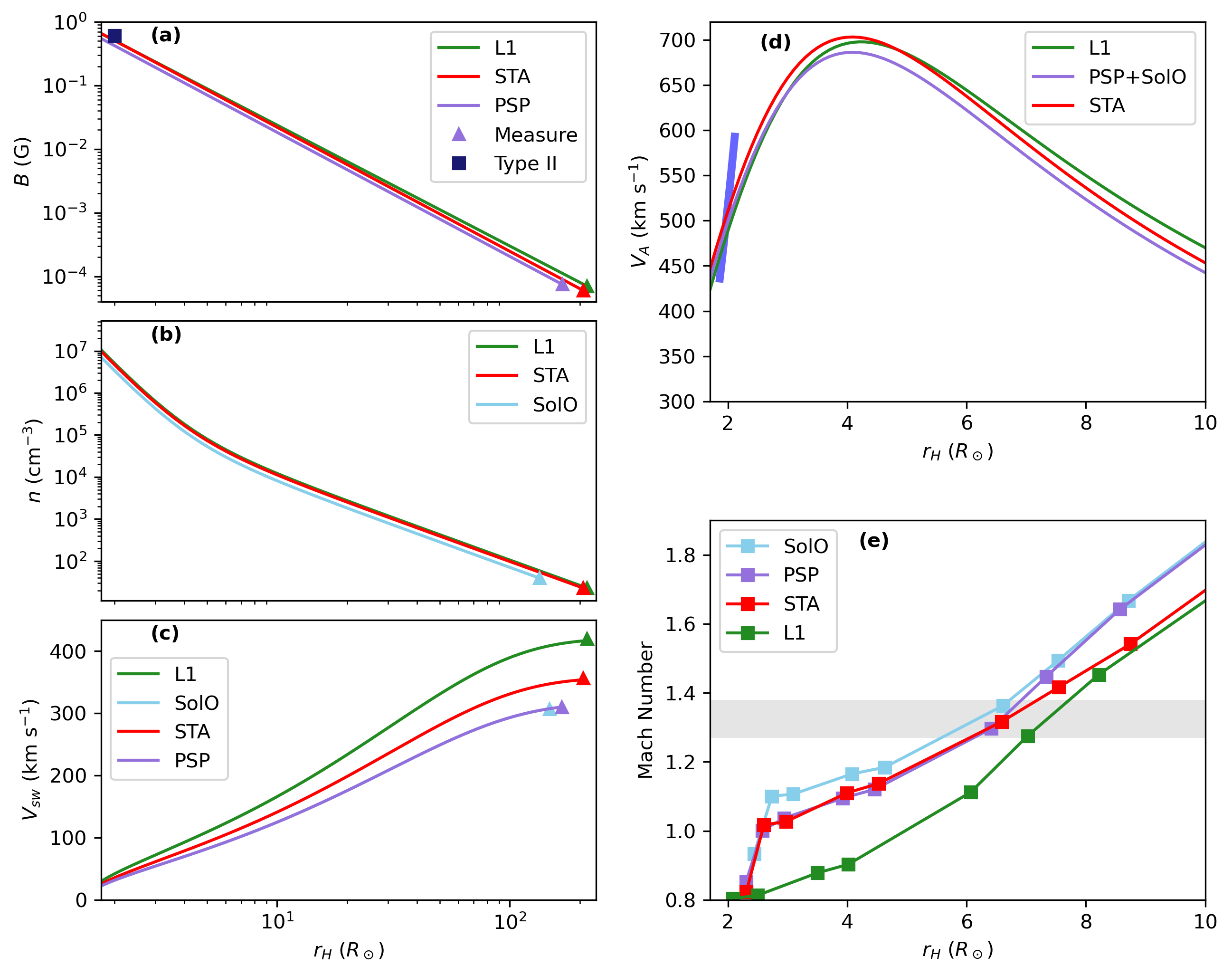}
    \caption{Variation of (a) magnetic field strength, (b) solar wind proton number density, (c) solar wind proton speed, (d) upstream Alfv\'{e}nic speed, and (e) shock Alfv\'{e}nic Mach number. In panel~(a), the filled square indicates the magnetic field strength at 2~$R_\odot$ based on the radio observations. In panel~(c), the curves of SolO and PSP are overlapped. The filled triangles in panels~(a)--(c) refer to the in-situ measurements. The blue thick bar in panel~(d) indicates the Alfv\'{e}n speed derived from the radio observations, and the shaded region in panel~(e) marks the estimated Mach numbers obtained by combining $\theta_{Bn}$ ranging between 10$^\circ$ and 45$^\circ$ and the DSA theory with $\gamma=3$.
    }
    \label{fig: mach}
\end{figure}
In Figure~\ref{fig: mach}, we estimate the evolution of the shock Alfv\'{e}nic Mach number and further explain the particle acceleration associated with the evolution of the shock. The Mach number is given as $M_A=(v_s-v_{sw}\cos\theta_{Bn})/v_A$, where $v_{sw}$ is the upstream solar wind speed, $v_A$ upstream Alfv\'{e}nic speed computed as $v_A = B/ \sqrt{\mu_0 \rho}$, where $B$ is the upstream magnetic field strength, $\mu_0$ is the magnetic permeability of free space, and $\rho$ is the plasma (mainly protons) density. We start from the in-situ measurements (averages within a six-hour duration before the SEP onset) (filled triangles) and track back the variation of the magnetic field strength, plasma number density, and solar wind speed (solid curve) as shown in panels~(a)--(c) of Figure~\ref{fig: mach}. The variation of the magnetic field strength along the heliocentric distance shown in panel~(a) is described by Equation~\ref{eq: b_var}:
\begin{equation} \label{eq: b_var}
    B=B_0r_H^{-\alpha},
\end{equation}
where $r_H$ is the heliocentric distance and $\alpha$ is the exponential factor which roughly equals 2. We note that there were no magnetic field measurements at SolO during the SEP event (PSP's measurements are used for SolO when the nominal magnetic field lines connecting the Sun and both spacecraft are closeby as described in Section~\ref{sec: sc_loc}). Here we adjust $\alpha$ (1.9 for L1, and 1.95 for PSP and STA) to make the extrapolated field strength consistent with that derived from the radio observations of 0.61~Gauss at 2~$R_\odot$ (filled square).

Panel~(b) of Figure~\ref{fig: mach} shows the variation of the plasma number density. Since Newkirk's model is only suitable for electron number density in the low corona, we use another model proposed by \citet{lablanc1998}. Under the condition that the proton and electron number densities are equal, the proton number density is thus given by
\begin{equation} \label{eq: leblanc}
    n(r_H) = n_0 [3.3\times 10^5 r_H ^{-2} + 4.1\times 10^6 r_H ^{-4} + 8 \times 10^7 r_H ^ {-6}] \ \rm cm^{-3}, 
\end{equation}
where $n_0$ is adjusted to make the model's estimation consistent with the in-situ measurements at SolO, STA, and at the spacecraft at L1 (i.e., Wind). There was no solar wind plasma data at PSP, and thus we assume PSP and SolO shared the same estimations. We note that the extrapolations using Leblanc's model from the in-situ measurements are generally consistent with Newkirk's model with $a=1$ in the low corona for this event (not shown here), which further validates using the one-fold Newkirk model in Section~\ref{sec: dis_radio}. Panel~(c) shows the variation of the solar wind speed by using the solar wind model of \citet{sheeley1997,sheeley1999}:
\begin{equation}\label{eq: vsw}
    v_{sw}(r_H)=v_{0}[1-e^{-\frac{r_H-r_0}{r_a}}],
\end{equation}
where $r_0=1.5$~$R_\odot$ denoting the heliocentric distance where $v_{sw}$ is zero, and $r_a$ indicating the distance at which the asymptotic solar wind speed is reached is set as 50~$R_\odot$.

The variation of $v_A$ is subsequently derived and shown in panel~(d) of Figure~\ref{fig: mach}: increasing from the solar surface to a peak of 700~km~s$^{-1}$ at the height of 4~$R_\odot$ and then decreasing. Such a profile of $v_A$ with a non-monotonic increase/decrease was also analyzed by \citet{evans2008} based on MHD simulations. The thick blue bar in panel~(d) indicates the $v_A$ value derived from the type II radio burst (see the two equivalent Equations~\ref{eq: va} and \ref{eq: va2} in Appendix~\ref{sec: app_radio}), which is overall consistent with the solid curves.

Combining the shock speed from the shock model at each cobpoint, we estimate the variation of the shock Alfv\'{e}nic Mach number by Equation~\ref{eq: va2} as shown in panel~(e) of Figure~\ref{fig: mach}. The Mach numbers of the four spacecraft are found to increase when the shock propagated in the middle to high corona. We note that the values at $\sim$2~$R_\odot$ are lower than the $\sim$1.19 result estimated from the type II radio emission observations (panel~(c) of Figure~\ref{fig: radio_model}). This is caused by the underestimation of the initial shock speed by the shock model 
, which is also shown in panel~(c) of Figure~\ref{fig: time}. 

In panel~(e) of Figure~\ref{fig: mach}, we shade a region by combining (a) $\theta_{Bn}$ roughly varying between 10$^\circ$ and 45$^\circ$ based on the shock model and nominal IMF lines at the cobpoints of the four spacecraft and (b) the diffusive shock acceleration \citep[DSA,][]{lee1983,lee2012} mechanism. DSA predicts that the spectra index ($\gamma$) of the SEP events based on the spectrum of ion differential intensity versus energy is only related to the shock compression ratio $X$, i.e., $\gamma \sim (X+2)/(2X-2)$. Considering $\gamma \sim 3$ at the four spacecraft for our event and Equation~\ref{eq: compres_r}, we can estimate (a) $X$ to be 1.6 and (b) $M_A$ as ranging between 1.27 and 1.38 (with $\theta_{Bn}$ ranging between $10^\circ$ and $45^\circ$ as described in Appendix~\ref{sec: app_radio}) corresponding to the shock strength to explain the measured $\gamma$. The heights of the shock at the cobpoints of the four spacecraft intersected in the shaded region range between 5.5--7.7~$R_\odot$. According to panel~(b) of Figure~\ref{fig: time}, the particles were released when the shock was at around 5--6~$R_\odot$, which is slightly lower than, but roughly consistent with, the heights estimated by using the DSA mechanism. We note that the release time of particles estimated at STA may be influenced by the in-situ magnetic field direction (see Section~\ref{sec: dis_uncer} below). The effects of inhomogeneities in the ambient medium leading to different particle properties are not discussed here \citep[e.g.,][]{kouloumvakos2022,kouloumvakos2023,wijsen2023}.

Overall, we find that the particle acceleration and release are primarily controlled by the evolution of the shock and upstream Alfv\'enic speeds. EUV waves are evidence for the presence of the fast-mode wave, but it can not guarantee that the shock is strong enough to accelerate particles at certain heights. For the SEP event studied here, even though the EUV wave intercepted the nominal field lines connected to each observer within a short time after the eruption, the shock was not strong enough to efficiently accelerate particles. During the outward propagation of the shock, its Mach number became larger due to (1) the increase in the shock speed as shown in panel~(c) of Figure~\ref{fig: time} \citep{kwon2017} and (2) the decrease in the background Alfv\'{e}n speed as shown in panel~(d) of Figure~\ref{fig: mach} \citep{kwon2018}. As a consequence, at around 5--6~$R_\odot$, we can deduce that the particles were efficiently accelerated and then released. We argue that the EUV wave did not contribute to the particle efficient acceleration and immediate release in this event as it was just the footprint of the shock on the coronal base.  The acceleration and release corresponded to the shock at higher altitudes when particle acceleration was more efficient. 

We further propose an explanation for past events where the EUV wave seemed to indicate the particle acceleration and release time \citep[e.g.,][]{rouillard2012, park2015}. These previous studies focused on SEP events observed by multiple spacecraft widely distributed in helio-longitude, and thus these events were very likely to be associated with a stronger and wider shock. For instance, the SEP events in the previous studies were mainly associated with halo CMEs with speeds $>$ 1000~km\,s$^{-1}$, whereas our CME is found to be a non-halo eruption (in both SOHO/LASCO and STEREO/COR2 FOVs) with a speed of $\sim$ 600~km\,s$^{-1}$. Stronger shocks would presumably be able to accelerate particles even at lower heights where EUV waves take place.

\subsection{Uncertainties in the Timing Comparison}\label{sec: dis_uncer}
In this section, we discuss some uncertainties in the timing estimation and illustrate that those uncertainties do not affect our major conclusion, i.e., that the particle release was delayed and such a delay was due to the evolution of the shock strength.

The first uncertainty is determining the observer's magnetic footpoint for the EUV wave connection. In some past studies \citep[e.g.,][]{park2013,park2015,lario2014,lario2017a}, apart from nominal Parker spiral field lines, the PFSS model was used to reconstruct the magnetic field in the low corona (i.e., $\le 2.5$~$R_\odot$) and thus had an additional estimate of the footpoints of the magnetic field lines connecting to each spacecraft. Taking advantage of the online magnetic connectivity tool \citep[\url{http://connect-tool.irap.omp.eu/new-home},][]{rouillard2020}, we found that the incorporation of the PFSS model can result in footpoints of the four spacecraft being even closer to the SEP source region compared to only using the assumption of nominal Parker spiral. This leads to an earlier connection between the EUV wave and magnetic footpoint and thus an even more significant delay in the particle release. Furthermore, based on the connectivity tool and an average EUV wave speed of 315~km\,s$^{-1}$, the uncertainties in the estimation of the time when the EUV wave intercepted the spacecraft nominal footpoints were tested to have no influence on the results of the delayed SEP release times for SolO, STA, and spacecraft at L1 (not shown here). However, different inputted magnetograms can have different outputs with even smaller uncertainties, which is outside the scope of this study. Besides, as shown in Figure~\ref{fig: time}, the estimated cobpoint heights have been around 2.5~$R_\odot$ very early in the event, and thus the PFSS model may not be suitable anymore because the coronal magnetic field has been already modified by the shock structure.

We further check whether our conclusions could be influenced by the effect of the heliospheric current sheet (HCS) structure on the location of the footpoints of the magnetic field lines connecting to the observing spacecraft \citep[e.g., see][]{badman2022}. We compare the magnetic polarities at the footpoints with those of in-situ observations. Figure~\ref{fig: psi_polarity} shows the distribution of the radial component of magnetic field ($B_r$) at 2.5~$R_\odot$ associated with a polarity inversion line (black line). The height of 2.5~$R_\odot$ refers to the PFSS height where magnetic field lines are forced to become radial \citep{badman2020}, and is consistent with the cobpoint heights at the times when the EUV wave intercepted the nominal footpoints of the four spacecraft (panel~(c) of Figure~\ref{fig: time}). This polarity map is based on the coronal solution of the Magnetohydrodynamic Algorithm outside a Sphere (MAS) model \citep[e.g.,][]{R2012} and made available online (\url{https://www.predsci.com/hmi/summary_plots.php}) by the Predictive Science Inc. group. The nominal Parker spiral field lines connecting to the spacecraft and passing through the height of 2.5~$R_\odot$ are determined (orange: SolO; black: PSP; red: STA, green: L1). Figure~\ref{fig: psi_polarity} shows that the nominal footpoints of SolO, PSP, and STA were located in the positive-polarity region (red area), whereas the L1 footpoint was located in the negative-polarity region (blue area) but very close to the positive-polarity region. It is consistent with the magnetic polarities measured in situ (Figure~\ref{fig: in-situ-para}): $B_r$ in the spacecraft-centered radial-tangential-normal (RTN) coordinate system measured at all spacecraft were positive at the SEP onset time. In addition, we also check the effect of a HCS within the same (positive) polarity region on the timing comparison. The intersections of the shock front based on the 3D shock model with the sphere of radius 2.5~$R_\odot$ for two time steps of 06:30~UT and 06:48~UT are overlaid with two dashed circles in Figure~\ref{fig: psi_polarity}. At 06:30~UT, the shock front is found to have already passed through all the footpoints and the majority of the positive-polarity region where the footpoints of PSP, SolO, and STA were positioned. At 06:48~UT, the region demarcated by this shock front includes negative-polarity regions, whereas all in-situ measurements at the four spacecraft suggest that $B_r$ at the magnetic footpoints should be positive. It indicates that the shock has reached the footpoints of all spacecraft before 06:48~UT (i.e., before the estimated particle release times), and thus the HCS crossing does not have a significant impact on the measurements studied in the paper.

\begin{figure}[!hbt]
    \centering
    \includegraphics[width=0.95\textwidth]{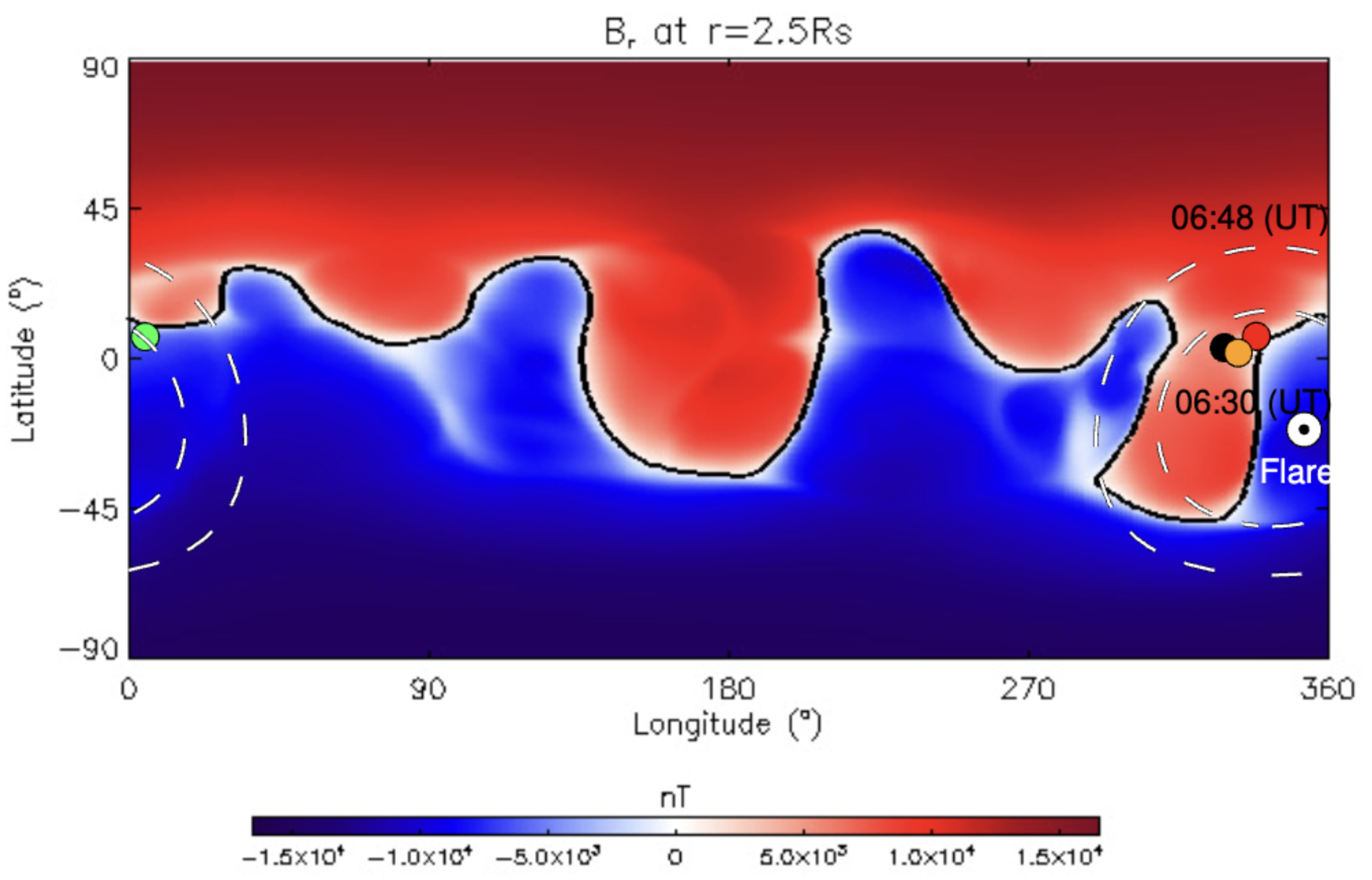}
    \caption{Magnetic polarity map of the radial component of the magnetic field ($B_r$) at the height of 2.5~$R_\odot$ for Carrington rotation 2249 (at 06:00~UT on 2021 September 28). This polarity map was taken from the summary plot on the Predictive Science Inc. website (\url{https://www.predsci.com/hmi/summary_plots.php}) based on the coronal solution of the MAS model. The black solid line positioned between the red and blue regions indicates the location of HCS. The flare site is shown by the filled white dot. The nominal Parker spiral lines connecting to the four spacecraft and passing through the surface at 2.5~$R_\odot$ are shown by four filled colored dots (orange: SolO; black: PSP; red: STA; green: L1). The two dashed circles indicate the intersections of the shock front based on the 3D shock model with the sphere of radius 2.5~$R_\odot$ at 06:30~UT (inner) and 06:48~UT (outer), respectively.
    }
    \label{fig: psi_polarity}
\end{figure}

The uncertainties in determining the particle release time include (a) the effects of the pre-event background intensity level and possible preceding SEP events on determining the particle intensity enhancement onset time, (b) the scattering effects undergone by the particles during their transport \citep[e.g.,][]{dresing2014,laitinen2015}, and (c) the influences of the interplanetary magnetic field and transients \citep[e.g.,][]{chhiber2021,lario2022}. As shown in Figure~\ref{fig: sep_obs}, the pre-event intensities were low before the SEP onset times and thus their effects are insignificant in evaluating the onset times of this SEP event, especially for particle intensities at high energies. Scattering conditions may result in a longer path length of particles and thus a delayed release time. Figure~\ref{fig: pitch_angle} shows anisotropy measurements at the onset of the SEP event at different spacecraft by plotting in the top panels the intensities measured along the different viewing angles of the selected particle detectors on board the four spacecraft. Using synchronized magnetic field data, we evaluate the pitch angles scanned by the central axis of each telescope (SolO/HET and PSP/LET) or the average of sector sides (STA/LET and Wind/3DP), which are plotted in the second panels of Figure~\ref{fig: pitch_angle}. The third and fourth panels show the magnetic field angles (elevation and azimuth) in RTN coordinates. Note that Wind/3DP particle intensities are binned in eight pitch angles scanned by the 3DP instrument on board this spinning spacecraft.

We find that at PSP and L1 (i.e., Wind), the anisotropies of the proton intensities at the onset of the event were large and the pitch angle distribution confirms that such large anisotropies are related to the particles coming from the Sun. SolO also observed strong anisotropies at the onset of the event, as intensities observed by the telescope pointing toward the Sun dominated over the intensities observed in other directions. However, the lack of magnetic field data does not allow us to evaluate the pitch angle of these particles. The strong anisotropies observed by SolO, PSP, and Wind suggest that the scattering effects for the particles registered at these three spacecraft were probably minor. We note that the anisotropy analysis at PSP was not affected by the spacecraft rotations. However, at STA, the observed anisotropy was small. In fact, both sides of the LET telescope detected pitch angles close to 90$^\circ$. Whereas the magnetic field had an strong out-of-ecliptic component (see Figure~\ref{fig: in-situ-para}), the orientation of the STA adopted since 2015 July 20 did not provide an appropriate configuration to detect particles propagating parallel to the nominal Parker spiral. Therefore, if the first particles arriving at STA propagated along the magnetic field, the LET telescope did not detect them. Consequently, estimates of the release times for those particles arriving at STA are uncertain.

We further use Figure~\ref{fig: time_delay} to simply test the influence of a longer path length on determining the particle release time. The estimate of a longer path length of particles can be caused by (a) scattering effects and (b) magnetic field line random walk under turbulent IMF conditions \citep{chhiber2021}. Since we cannot quantitatively measure the path lengths affected by these two factors for particles at different energies, the assumption is that those particles propagate along the same path length. The shaded regions in the figure show the onset time of protons at different energies by assuming (1) the particles being released when the EUV wave connected to the magnetic footpoint and (2) a range of path length increasing from 1 to 1.4 times the nominal spiral length. It can be clearly seen that the increased field length can be consistent between the particle release time and the EUV wave connection time at low ($<$1--2~MeV) energies, but it is not consistent with particles at higher energies (the real onset time at the four spacecraft was delayed). In addition, Figure~\ref{fig: in-situ-para} shows that there were no large-scale interplanetary transients during the onset phase of the SEP event at the four spacecraft that could have been responsible for longer path lengths.

\section{Summary and Conclusions}\label{sec: sum}
The temporal relationship between the release of energetic particles and the connection of the EUV wave to the spacecraft magnetic footpoint has been debated for decades. Taking advantage of the measurements taken by SolO and PSP close to the Sun and by STA and L1 spacecraft at different longitudes, together with a rare configuration where PSP, STA and SolO were nominally connected to closeby regions of the Sun, we studied the temporal correlation between SEP release and EUV wave connection for a SEP event on 2021 September 28. During this time, SolO, PSP, and STA shared similar nominal magnetic footpoints but were at different heliocentric distances. We find that for the four spacecraft, the particle release time was delayed compared to the time when the EUV wave connected to the spacecraft nominal footpoint by around 30 to 60 minutes. Such a delay was significant even when considering a variety of uncertainties.

We then combined a geometrical shock model based on remote-sensing observations from multiple viewpoints and type II radio burst observations to investigate the evolution of the shock properties. We find that the shock became stronger during its propagation from the low to middle corona. Based on the diffusive shock acceleration mechanism and the event-integrated energy spectrum, we estimate shock properties that are consistent with the results of the shock model at the particle release time. This provides a scenario to explain the delay for this event: the shock could not be strong enough to efficiently accelerate energetic particles, especially in the low solar corona where the shock propagated in the form of EUV waves; efficient acceleration occurred after the shock propagated to the high corona, e.g., $\sim$5--6~$R_\odot$,  and experienced strength enhancement, which explains the delayed particle release time as compared to when the EUV wave passed the footpoint of the spacecraft on the solar surface. 

\begin{acknowledgments}
We thank the data from SolO, PSP, STEREO, Wind, SOHO, SDO, and the e-CALLISTO Australia/ASSA station. Solar Orbiter is a space mission of international collaboration between ESA and NASA, operated by ESA. Parker Solar Probe was designed, built, and is now operated by the Johns Hopkins University Applied Physics Laboratory as part of NASA’s LWS program. We also thank the Predictive Science Inc. group for making available the magnetic polarity map used in this study on their website (\url{https://www.predsci.com/hmi/summary_plots.php}). B.Z. and N.L. acknowledge NASA grants 80NSSC19K0831 and 80NSSC20K0431 and the NSF grant AGS-2301382. D.L. acknowledges support from NASA Living With a Star (LWS) programs NNH17ZDA001N-LWS and NNH19ZDA001N-LWS, the Goddard Space Flight Center Internal Scientist Funding Model (competitive work package) program, and the Heliophysics Innovation Fund (HIF) program. N.C. acknowledges funding support from CNES and from the Initiative Physique des Infinis (IPI), a research training program of the Idex SUPER at Sorbonne Universit\'{e}. R.-Y.K acknowledges support by basic research funding from the Korea Astronomy and Space Science Institute (KASI2023185007).
\end{acknowledgments}

\begin{figure}[!hbt]
    \centering
    \includegraphics[width=0.8\textwidth]{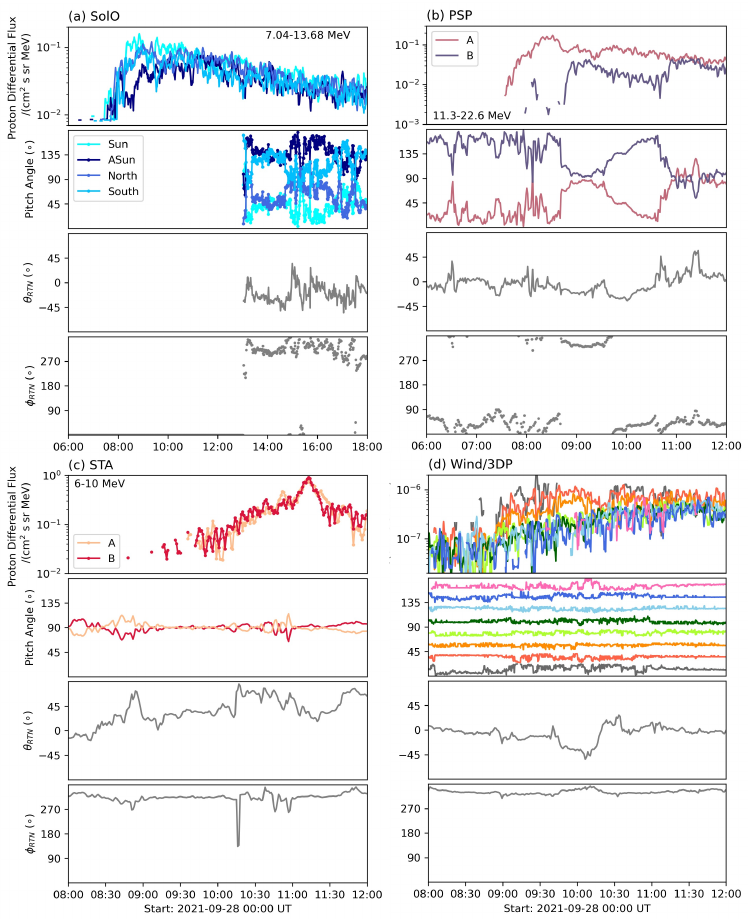}
    \caption{Anisotropies of proton intensities at the SEP onset observed by the selected particle detectors along different viewing angles on board the four spacecraft (top panels) associated with the pitch-angle distribution (second panels) and elevation angle ($\theta_{RTN}$) and azimuth angle ($\phi_{RTN}$) of the IMF in RTN coordinates (third and fourth panels) at (a) SolO/HET, (b) PSP/LET, (c) STA/LET, and (d) Wind/3DP. At SolO, the four colors are related to the sunward, anti-sunward, north, and south apertures of SolO/HET. At PSP, the two colors are related to the LET A and B heads. At STA, the two colors are related to the measurements of the two average sector sides (A and B). At Wind, the eight colors are related to different measurements along eight pitch-angles as shown in the second row of panel~(d).
    }
    \label{fig: pitch_angle}
\end{figure}

\begin{figure}[!hbt]
    \centering
    \includegraphics[width=0.9\textwidth]{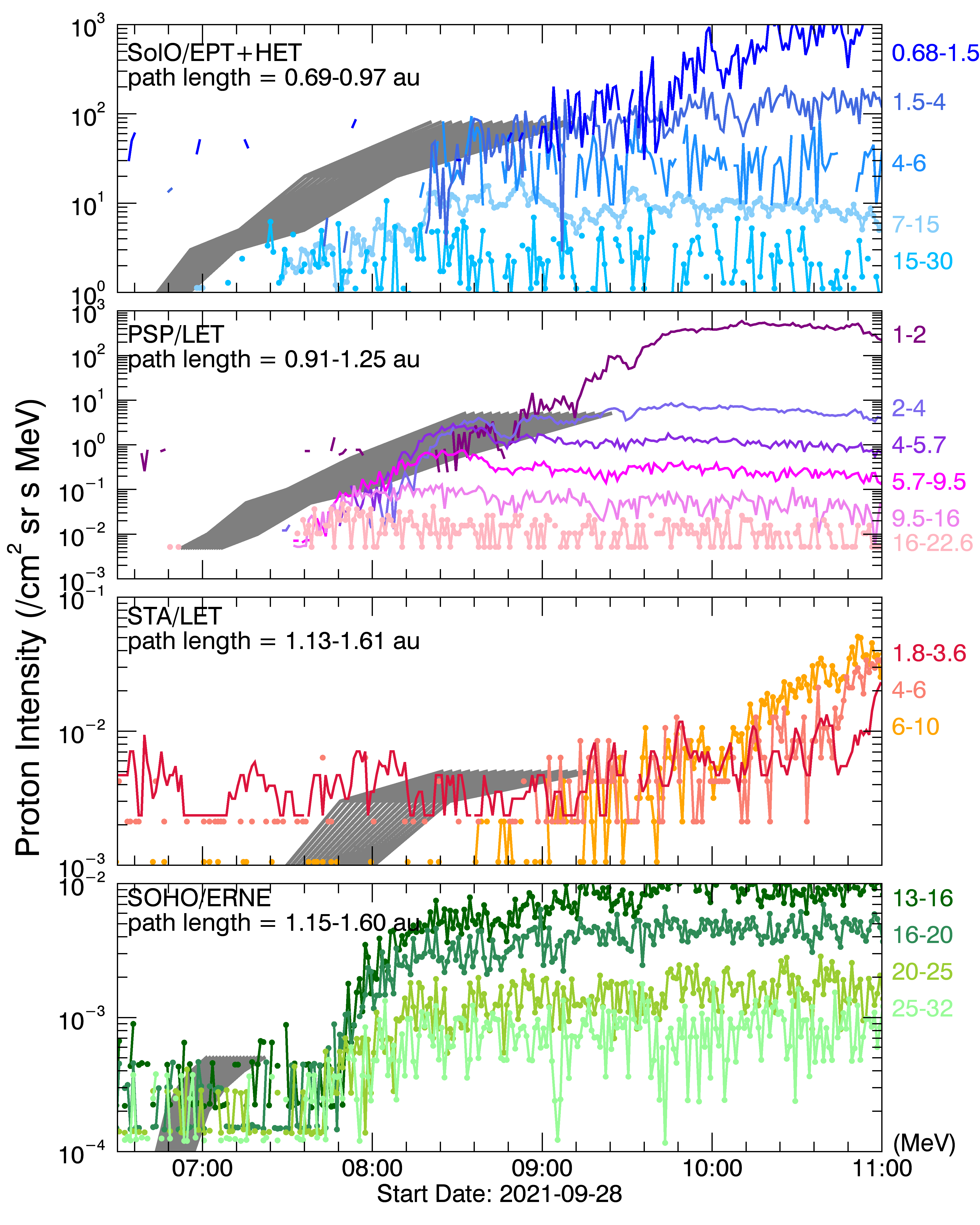}
    \caption{Test of the onset time of energetic particles at different energies by assuming the particles were released when the EUV wave connected to the spacecraft magnetic footpoint and transported along a longer path length. The onset times under the assumptions of different path lengths are shown by the shaded region. One can see more details in the main text.
    }
    \label{fig: time_delay}
\end{figure}

\clearpage

\begin{sidewaystable}[!htb]
	\centering
	\caption{Table of the spacecraft, solar wind, IMF, and SEP properties.}
	\label{tab: sep_info}
	\begin{tabular}{|c|c|c|c|c|c|c|c|c|c|c|c|}
		\hline
        \multirow{2}{*}{S/C} & \multirow{2}{*}{Inst.} & $r_H$ & $\phi_C$ & $\theta_C$ & $v_{\rm sw}$ & $\phi_{\rm F}$ & $l$ & Energy & $t_{\rm onset}$ & $t_{\rm SPR}$ & $t_{\rm EUV}$ \\
        
        & & (au) & ($^\circ$) & ($^\circ$) & (km~s$^{-1}$) & ($^\circ$) & (au) & (MeV) & (2021-09-28) & (2021-09-28) & (2021-09-28) \\ \hline
        
        \multirow{2}{*}{SolO} & \multirow{2}{*}{HET} & \multirow{2}{*}{0.62} & \multirow{2}{*}{285} & \multirow{2}{*}{2.0} & \multirow{2}{*}{307} & \multirow{2}{*}{334} & \multirow{2}{*}{0.69} & 7.0--13.7 & 07:32~UT ($\pm 5$~min) & 06:52~UT ($\pm 5$~min) & \multirow{2}{*}{06:25~UT} \\
        & & & & & & & & 13.7--25.1 & 07:21~UT ($\pm 3$~min) & 06:52~UT ($\pm 3$~min) & \\ \hline
        
        \multirow{2}{*}{PSP} & \multirow{2}{*}{LET} & \multirow{2}{*}{0.78} & \multirow{2}{*}{268} & \multirow{2}{*}{3.5} & \multirow{2}{*}{310} & \multirow{2}{*}{330} & \multirow{2}{*}{0.91} & 6.7--13.5 & 07:37~UT ($\pm 3$~min) & 06:44~UT ($\pm 2$~min) & \multirow{2}{*}{06:23~UT} \\
        & & & & & & & & 13.5--26.9 & 07:33~UT ($\pm 3$~min) & 06:55~UT ($\pm 3$~min) & \\ \hline
        
        \multirow{2}{*}{STA} & \multirow{2}{*}{LET} & \multirow{2}{*}{0.96} & \multirow{2}{*}{272} & \multirow{2}{*}{6.9} & \multirow{2}{*}{357} & \multirow{2}{*}{337} & \multirow{2}{*}{1.13} & \multirow{2}{*}{6--10} & \multirow{2}{*}{08:30~UT ($\pm 3$~min)} & \multirow{2}{*}{07:16~UT ($\pm 3$~min)} & \multirow{2}{*}{06:24~UT} \\
        & & & & & & & & & & & \\ \hline
        
        \multirow{2}{*}{L1} & \multirow{2}{*}{ERNE} & \multirow{2}{*}{1.00} & \multirow{2}{*}{312} & \multirow{2}{*}{6.8} & \multirow{2}{*}{420} & \multirow{2}{*}{10} & \multirow{2}{*}{1.15} & 13--20 & 07:48~UT ($\pm 4$~min) & 06:55~UT ($\pm 4$~min) & \multirow{2}{*}{06:20~UT} \\
        & & & & & & & & 20--32 & 07:43~UT ($\pm 6$~min) & 07:01~UT ($\pm 6$~min) & \\ \hline
        
	\end{tabular}
	
	\setlength{\extrarowheight}{10pt}
	\hspace*{\fill} \\
	\footnotesize{The table lists the spacecraft and instruments, the heliocentric distance ($r_H$), Carrington longitude ($\phi_C$), and Carrington latitude ($\theta$) of the spacecraft, the in-situ solar wind proton speed ($v_{\rm sw}$), nominal magnetic footpoint ($\phi_F$) in Carrington longitude and IMF line length ($l$) between the SEP source region and spacecraft, the energy ranges used for the onset time identification, the identified onset time ($t_{\rm onset}$), the particle release time ($t_{\rm SPR}$) in the related energy range, and the time when the EUV wave connected to the nominal magnetic footpoint of the spacecraft ($t_{\rm EUV}$).
	}
\end{sidewaystable}
\clearpage

\begin{appendix}

\setcounter{figure}{0}
\renewcommand{\thefigure}{A\arabic{figure}}

\section{In-situ Measurements}\label{sec: app_insitu}
\begin{figure}[!hbt]
    \centering
    \includegraphics[width=0.95\textwidth]{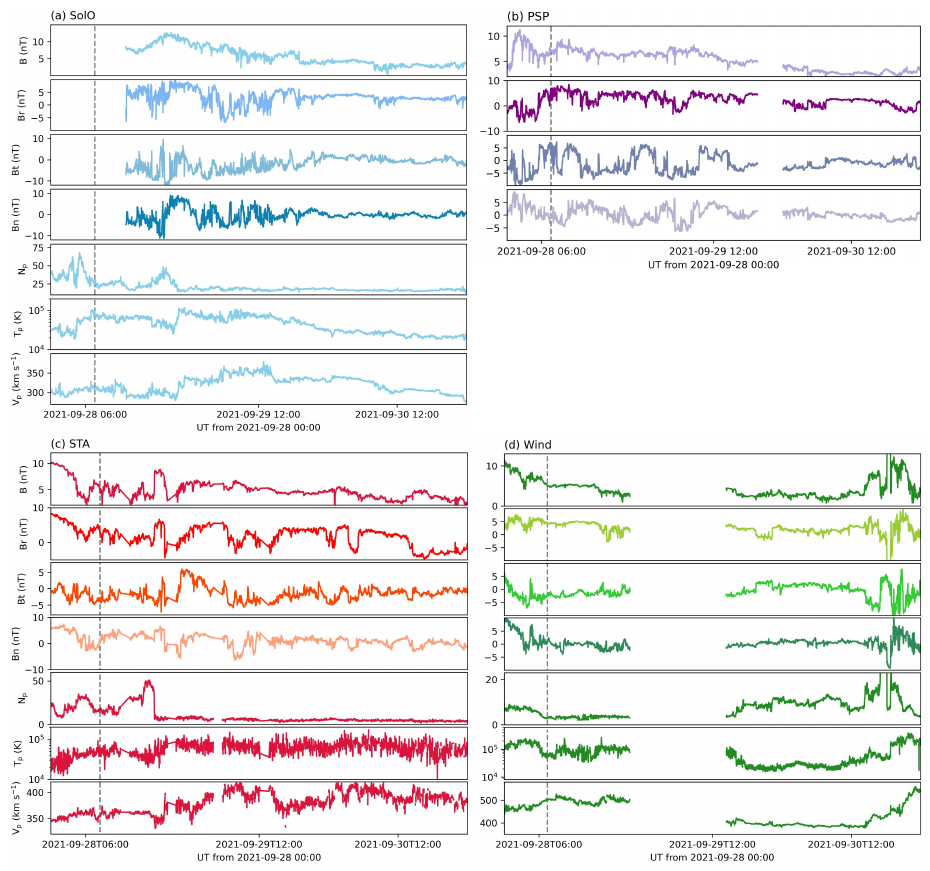}
    \caption{In-situ IMF (in RTN coordinates) and solar wind plasma observations of 1-minute resolution at (a) SolO, (b) PSP (not including plasma measurements), (c) STA, and (d) Wind during September 28 and 30. The vertical dashed lines indicate the onset time of the proton flux enhancements.
    }
    \label{fig: in-situ-para}
\end{figure}
Figure~\ref{fig: in-situ-para} shows the in-situ measurements of the IMF (in RTN coordinates) and solar wind plasma number density, temperature, and speed at SolO (a), PSP (b), STA (c), and Wind (d) during September 28 and 30, respectively. The vertical dashed line indicates the SEP onset time. We calculate the average IMF strengths, solar wind proton number densities, and/or speeds at the four spacecraft within a six-hour duration before the SEP onset. The estimated speeds are listed in Table~\ref{tab: sep_info}, and the IMF and proton number density results are used in Figure~\ref{fig: mach}. We note that there were no solar wind measurements with good-quality flags at PSP during the event, and thus we assume the associated solar wind speed as 310~km~s$^{-1}$ based on the close locations between SolO and PSP. We also find that during the onset phase of the SEP event, there were no clear signatures of large-scale transients detected in situ at the four spacecraft. 

\section{Model of Split-Band Structure in Type II Radio Bursts} \label{sec: app_radio}
In this section, we describe the upstream-downstream model of the split-band structure in type II radio bursts, in which the band splitting is assumed to be due to the emissions from the upstream (ahead) and downstream (behind) regions of the shock front \citep{smerd1975,vrsnak2001,vrsnak2002}. The radio emission frequency and electron number density ($n_e$, in units of cm$^{-3}$) are related by
\begin{equation} \label{eq: freq_n}
    f = 9 \times 10^{-3} n_e^{1/2} \ \rm MHz.    
\end{equation}
The density compression ratio ($X$) can be expressed as
\begin{equation} \label{eq: compres_r}
    X=\frac{n_{e2}}{n_{e1}}=\frac{f_u^2}{f_l^2}=(1+\frac{\Delta f}{f_l})^2,
\end{equation}
where $n_{e1}$ and $n_{e2}$ are the upstream and downstream number densities, $\Delta f = f_u - f_l$, and $f_l$ and $f_u$ are the upstream and downstream frequencies, respectively. Based on the Rankine-Hugoniot jump conditions, the relationship between $X$ and the upstream Alfv\'{e}nic Mach number ($M_A$) is described by
\begin{equation} \label{eq: compres_r}
    M_A ^2 = (M_{A\perp} \sin \theta_{Bn})^2+(M_{A\parallel} \cos \theta_{Bn})^2,
\end{equation}
where $M_{A\perp}=\sqrt{\frac{X(5+X)}{2(4-X)}}$ and $M_{A\parallel}=\sqrt{X}$ based on the assumptions of (a) the plasma-to-magnetic pressure ratio $\beta \to 0$ and (b) the adiabatic index to be $5/3$, and $\theta_{Bn}$ is the angle between the magnetic field and shock normal. In the case of a perpendicular shock ($\theta_{Bn}=90^\circ$), Equation~\ref{eq: compres_r} becomes:
\begin{equation} \label{eq: compres_r_perpen}
    M_A = \sqrt{\frac{X(5+X)}{2(4-X)}}.
\end{equation}
One can refer to \citet{vrsnak2002} and \citet{kwon2018} for detailed descriptions of the $X$-$M_A$ relationship. We note that the relationship of $M_A=\sqrt{\frac{3X}{4-X}}$ as shown in \citet{smerd1975} was based on the hydrodynamic assumption. The source region of the type II radio burst cannot be identified during this SEP event. Thus, considering the angular separations of 10$^\circ$--45$^\circ$ between the nominal Parker spiral lines and the shock normal at the cobpoints of the four spacecraft used in this paper (also see the right columns of Figure~\ref{fig: shock}), we use an average value of $\theta_{Bn}\sim 20^\circ$ (larger $\theta_{Bn}$ leads to a higher $M_A$ with a fixed $X$ which is larger than 1) for a simple estimation.

Combining the frequency drift with time ($df_l/dt$) and the coronal electron number density model, we can estimate the shock speed $v_s$. Following \citet{smerd1975}, we use the Newkirk coronal model \citep{newkirk1961}, in which $n_e = 4.2a \times 10^{4+4.32 \frac{R_\odot}{r_H}}$, where $r_H$ is the heliocentric distance and $a$ is set to be 1 (i.e., the one-fold Newkirk model; the effectiveness of $a=1$ is discussed in Section~\ref{sec: dis_acc}).
The shock speed can be described as:
\begin{equation} \label{eq: vshock}
    v_{s} = \frac{6.529 \times 10^5 |df_l/dt|}{f_l [\log _{10} (f_l / a^{1/2}) - 0.415^2]} \ \rm km~s^{-1}.
\end{equation}
The upstream Afv\'{e}nic speed is estimated as
\begin{equation} \label{eq: va}
    v_{A} = B/\sqrt{\mu_0 \rho}=1.96\times 10^4 B / f_l \ \rm km~s^{-1},
\end{equation}
where $B$ is the upstream magnetic field strength in Gauss, $\mu_0$ is the magnetic permeability of free space, and $\rho$ is the plasma (mainly protons) density. Therefore, if $v_s$ and $M_A$ are estimated, the upstream $B$ can also be derived. We have
\begin{equation} \label{eq: va2}
    v_{A} = (v_s-v_{sw} \cos\theta_{Bn})/M_A,
\end{equation}
where $v_{sw}$ is the upstream solar wind speed (based on Equation~\ref{eq: vsw}, $v_{sw}$ is estimated as $\sim$40~km\,s$^{-1}$ during the event). In previous studies which focused on the shock in the low corona, the solar wind speed can be neglected. However, in this study, we extend our estimation of $M_A$ up to $\sim$9~$R_\odot$ (panel~(e) of Figure~\ref{fig: mach}), and $v_{sw}$ needs to be considered. Finally, $B$ can be expressed as:
\begin{equation} \label{eq: bmag}
    B \ ({\rm G}) = 5.1\times 10^{-5} (v_s-v_{sw}\cos\theta_{Bn}) f_l/M_A.
\end{equation}

\end{appendix}

\bibliography{sample631}{}
\bibliographystyle{aasjournal}

\end{document}